\RequirePackage{graphicx}
\documentclass[prd, superscriptaddress, nofootinbib, floatfix,preprintnumbers]{revtex4-2}   

\usepackage{tikz}
\usepackage{xcolor}
\usepackage[subfigure]{graphfig}
\usepackage[normalem]{ulem} %to strike the words
\usepackage{wrapfig}
\usepackage{amsmath}
\usepackage{amsfonts}
\usepackage{amssymb}
\usepackage{booktabs}
\usepackage{multirow}
\usepackage{slashed}
\usepackage{physics}
\usepackage{tabularx}
\usepackage{soul}
\usepackage{ulem}
\usepackage{hhline}
\usepackage{float}
\usepackage[colorlinks=true, linkcolor=blue, citecolor=blue, urlcolor=blue]{hyperref}
\usepackage{longtable}

\usepackage{nccmath}
\usepackage{esvect}
\usepackage{slashed}

\newcommand\befs{\begin{figure*}}
\newcommand\eefs[1]{\label{fig:#1}\end{figure*}}
\newcommand\bef{\begin{figure}}
\newcommand\eef[1]{\label{fig:#1}\end{figure}}
\newcommand\beq{\begin{equation}}
\newcommand\eeq[1]{\label{#1}\end{equation}}
\newcommand\beqa{\begin{eqnarray}}
\newcommand\eeqa[1]{\label{#1}\end{eqnarray}}
\newcommand\bet{\begin{table}}
\newcommand\eet[1]{\label{tb:#1}\end{table}}
\newcommand\bets{\begin{table*}}
\newcommand\eets[1]{\label{tb:#1}\end{table*}}

\def\be{\begin{equation}}
\def\ee{\end{equation}}
\newcommand{\bea}{\begin{eqnarray}}
\newcommand{\eea}{\end{eqnarray}}
\newcommand{\ba}{\begin{align}}
\newcommand{\ea}{\end{align}}

\newcommand{\nn}{\nonumber}

\begin{document}

\title{Elastic and resonance structures of the nucleon from hadronic tensor in lattice QCD: implications for neutrino-nucleon scattering and hadron physics} 

\newcommand*{\SCNUa}{Key Laboratory of Atomic and Subatomic Structure and Quantum Control (MOE), Guangdong Basic Research Center of Excellence for Structure and Fundamental Interactions of Matter, Institute of Quantum Matter, South China Normal University, Guangzhou 510006, China}
\newcommand*{\SCNUb}{Guangdong-Hong Kong Joint Laboratory of Quantum Matter, Guangdong Provincial Key Laboratory of Nuclear Science, Southern Nuclear Science Computing Center, South China Normal University, Guangzhou 510006, China }
\affiliation{\SCNUa} 
\affiliation{\SCNUb} 

\newcommand{\NMSU}{Department of Physics, New Mexico State University, Las Cruces, NM 88003, USA.}\affiliation{\NMSU}
     
\newcommand*{\RBRC}{RIKEN-BNL Research Center, Brookhaven National Laboratory, Upton, NY 11973, USA.}\affiliation{\RBRC}

\newcommand*{\BNL}{Physics Department, Brookhaven National Laboratory, Upton, NY 11973, USA.}\affiliation{\BNL}

\newcommand*{\UKY}{Department  of  Physics  and  Astronomy,  University  of  Kentucky,  Lexington,  KY  40506, USA.}\affiliation{\UKY} 

\newcommand*{\DU}{Department of Theoretical Physics, University of Dhaka, Dhaka 1000, Bangladesh.}\affiliation{\DU}

\newcommand*{\CAS}{CAS  Key  Laboratory  of  Theoretical  Physics,  Institute  of  Theoretical  Physics, Chinese  Academy  of  Sciences,  Beijing  100190,  China.}\affiliation{\CAS}

\newcommand*{\LBL}{Nuclear Science Division, Lawrence Berkeley National Laboratory, Berkeley, CA 94720, USA.}\affiliation{\LBL}

%%%%%%%%%%%%%%%%%%%%%%%%%%%
% AUTH BLOCK
%%%%%%%%%%%%%%%%%%%%%%%%%%%
\author{Jian Liang}\email{jianliang@scnu.edu.cn}
\affiliation{\SCNUa} 
\affiliation{\SCNUb} 
\author{Raza Sabbir Sufian}\email{gluon2025@gmail.com}\affiliation{\NMSU}\affiliation{\RBRC}\affiliation{\BNL}
\author{Bigeng Wang}\email{bwa271@g.uky.edu}\affiliation{\UKY}
\author{\mbox{Terrence Draper}}\affiliation{\UKY}
\author{Tanjib Khan}\affiliation{\DU}
\author{Keh-Fei Liu}\email{liu@g.uky.edu}\affiliation{\UKY}
\author{Yi-Bo Yang}\affiliation{\CAS}
\author{Christian Zimmermann}\affiliation{\UKY}\affiliation{\LBL}

\vspace*{-2.5cm}
\begin{center}
\large{
\vspace*{0.4cm}
\includegraphics[scale=0.20]{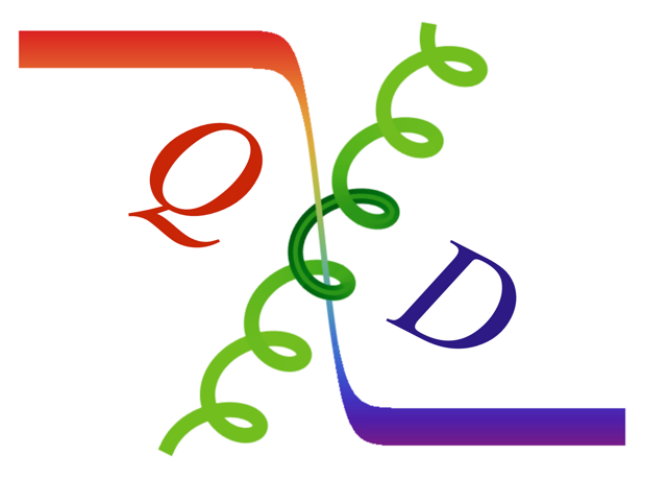}\\
\vspace*{0.4cm}
($\chi$QCD  Collaboration)
}
\end{center}

\begin{abstract}
%%%%%
 We compute the Euclidean hadronic tensor from  charge density operators and extract elastic and resonance structures by employing exponential fits to the four-point function correlator, as well as a Bayesian reconstruction inverse algorithm to obtain the corresponding spectral density for qualitative comparison. We present the determination of the nucleon's Sachs electric form factor using the hadronic tensor formalism and verify that it is consistent with that from the conventional three-point function calculation.
Beyond the elastic peak, we observe a structure located approximately $0.5-0.7$ GeV above the nucleon mass in the Bayesian reconstruction. This structure is interpreted as a mixture of the Roper resonance ($N(1440)$), and  states with both positive and negative parities in this mass region,  as well as multi-hadron states.  Assuming the observed structure is dominated by  $J^P = 1/2^{\pm}$
 states, we extract the transition electric form factor $G_E^*(Q^2)$ and the corresponding longitudinal helicity amplitude $S_{1/2} (Q^2)$, and compare them with those  determined from the CLAS experimental data of  nucleon-to-Roper transition.
Although  fitting to the four-point correlation function or using the inverse algorithm does not resolve individual resonances, it nevertheless enables the determination of total inclusive lepton–nucleon scattering cross sections in appropriate energy bins. This lattice QCD calculation presents the first major step toward studying the inclusive $N\to X$ contributions within the hadronic tensor formalism.

\end{abstract}

\maketitle

%%%%%

\newpage
\section{Introduction} \label{sec:intro1}

Understanding the hadronic tensor occupies a central and indispensable role in the domains of particle and nuclear physics through its relevance in quasi-elastic, resonance, and deep inelastic scattering experiments. A profound knowledge of hadronic tensors is essential to discern valuable information about the internal structure of hadrons and the distribution of quarks and gluons within hadrons, and to gain insights into the strong  interactions. Lattice quantum chromodynamics (QCD) determinations of the hadronic tensor~\cite{Liu:1993cv, Liu:1999ak} offer an opportunity to significantly influence our comprehension of the neutrino-nucleon interactions and can provide valuable complementary theoretical constraints while interpreting data from  the neutrino oscillation experiments. Furthermore, understanding the excitation spectrum of baryons and mesons, as well as characterizing the effective degrees of freedom in various QCD coupling regimes, stand as some of the most important and undeniably challenging endeavors within the field of hadronic physics where the hadronic tensor can provide direct information on these transition spectra and spectral weights. The potential of this approach was promptly acknowledged in the white papers relevant to neutrino physics~\cite{Kronfeld:2019nfb,Ruso:2022qes}. It was highlighted that the scattering cross-section for neutrino-nucleon ($\nu-N$) interactions can be obtained by utilizing both the leptonic and hadronic tensors, with the nonperturbative hadronic tensor computable through  the first-principles lattice QCD calculations.

The investigation of the hadronic tensor within lattice QCD will offer a significant influence on our comprehension of the findings from the Deep Underground Neutrino Experiment (DUNE)~\cite{DUNE:2020lwj} and Hyper-K~\cite{Hyper-Kamiokande:2018ofw} experiments. Neutrino oscillation experiments are performed using scattering on nuclear targets, and it is crucial to construct appropriate models for neutrino-nucleus ($\nu-A$) interactions, which involve modifications to $\nu-N$ interactions due to nuclear effects. These model buildings are essential to effectively analyze the data and accurately reproduce and interpret the observed cross sections, enabling the investigation of neutrino oscillations. This task requires the direct incorporation of scattering amplitudes at the nucleon level. Gaining precise knowledge of the $\nu-N$ scattering cross-section within the few-GeV range is indispensable for exploring charge-parity (CP) symmetry violation in the leptonic sector. Furthermore, the total cross sections for $\nu-N$ or $\nu-A$ scattering at various neutrino beam energies are due to  contributions from quasi-elastic (QE), resonance (RES), shallow inelastic (SIS), and deep inelastic scattering (DIS) processes, as discussed in \cite{Formaggio:2012cpf}. At energies below $1$ GeV, in which a nucleon is knocked out of the target nucleus, QE scattering ($\nu_l + n \to l^- + p$, $\overline{\nu}_l + p \to l^+ + n$) primarily governs the scattering cross-section, which can be  derived using the hadronic tensor formalism. On the contrary, within the RES, SIS, and DIS energy ranges, the cross section incorporates inputs from inelastic neutrino-nucleon interactions. In the region of resonant pion production, the contribution of the $\Delta$-resonance to the neutrino cross section is the most prominent one. However, the contributions coming from the second resonance region, which includes three isospin-$\frac{1}{2}$ states: $D_{13}(1520),~ P_{11}(1440)$ and $S_{11}(1535)$, are also important. 

 This necessitates the determination of the hadronic tensor to encompass all-inclusive contributions, information that can be obtained from lattice QCD calculations. However, grasping the entirety of the resonant region, which encompasses a spectrum of multi-particle states, poses a formidable challenge. Yet, there is promise in reconstructing these spectral functions through the use of two currents in different channels, enabling the computation of the inclusive hadronic tensor. Furthermore, the SIS region, situated between resonances and deep inelastic scattering, currently lacks a consistent theoretical model for its description~\cite{NuSTEC:2017hzk}. It is likely that future calculations of the hadronic tensor in the SIS region will offer valuable theoretical insights, as suggested in \cite{Kronfeld:2019nfb}. The desired precision for these measurements, encompassing systematic uncertainties (which presently dominate the pursuit of physics discoveries), lies within the range of $1- 4\%$ for the $\nu_\mu \to \nu_e$ and $\bar{\nu}_\mu \to \bar{\nu}_e$ oscillation probabilities~\cite{NuSTEC:2017hzk,Kronfeld:2019nfb,Ruso:2022qes}. Hence, the imperative of directly determining the hadronic tensor through lattice QCD becomes essential to fully harness the potential of new neutrino experimental facilities such as DUNE and Hyper-K.

For instance, in~\cite{Sufian:2018qtw}, lattice QCD inputs of the strange quark form factors and strange quark contribution to nucleon spin from~\cite{Sufian:2016pex,Sufian:2017osl,Sufian:2016vso,Liang:2018pis} were utilized. These were then combined with experimental measurements of neutrino-nucleon scattering differential cross sections~\cite{MiniBooNE:2010xqw,MiniBooNE:2013dds}, without considering nuclear effects, in order to determine the neutral current weak axial form factor that remains largely unknown. Conversely, the ultimate objective is to directly determine nucleon-level structure functions crucial for neutrino scattering cross sections from lattice QCD. In this regard, in an ideal scenario where all systematic uncertainties are well-controlled, the computation of the hadronic tensor through lattice QCD can provide predictions for nucleon-level  structure functions, encompassing various types of current combinations, such as vector-vector, vector-axial, and axial-axial currents.

 The formalism of the hadronic tensor also facilitates the exploration of nucleon resonance excitations that manifest through intermediate processes like $\gamma N \to N^*$. Consequently, this approach enables the determination of cross sections in terms of electromagnetic transition form factors or helicity transition amplitudes. Such an application of the hadronic tensor offers useful information to investigate baryon resonances via electromagnetic excitations~\cite{Burkert:2005ak} and the exclusive electroproduction of mesons involving baryon resonances~\cite{Aznauryan:2012ba} which is directly related to the experimental program of the determination of generalized parton distributions (GPDs) at Jefferson Lab. While the hadronic tensor formalism in lattice QCD  allows to determination GPDs, they also provide complementary information in the GPD program through their access to the resonance transition form factors. The investigation of nucleon resonances stands as a significant objective for the CEBAF Large Acceptance Spectrometer (CLAS) at Jefferson Lab, as well as for lattice QCD spectroscopy programs~\cite{Detmold:2019ghl}. Notably, a comprehensive understanding of the nucleon resonance excitation within electromagnetic interactions yields crucial insights into the dynamics of strong interactions within the realm of quark confinement~\cite{Aznauryan:2011qj}. In addition, the determination of the transition amplitudes in a wide range of momentum transfer allows to map out the quark transverse charge distributions that induce these transitions~\cite{Carlson:2007xd,Tiator:2008kd,Aznauryan:2011qj}. Notably, the hadronic tensor formalism~\cite{Liu:1993cv,Liu:1999ak} also allows  one to determine the  $x$-dependent structure functions, being complementary to other lattice QCD formalisms proposed in~\cite{Braun:2018brg,Ji:2013dva,Ji:2014gla, Chambers:2017dov,Radyushkin:2017cyf,Ma:2017pxb}.

It is worth mentioning that the initial proposals of the hadronic tensor~\cite{Liu:2020okp,Liu:1998um} allowed for the study of  both the connected and disconnected sea partons, thereby realizing the violation of the Gottfried sum rule~\cite{Gottfried:1967kk} in~\cite{Liu:1993cv,Hou:2022ajg}. Additionally, by numerically computing different topologies of the quark lines in the four-point current-current correlator, the hadronic tensor enables the direct estimation of contributions from higher-twist ``cat's ears" diagrams, as well as the leading-  and higher-twist ``hand-bag" diagram in the DIS processes.  Calculations of hadron electromagnetic  polarizabilities using the four-point hadronic tensor calculation can be found in~\cite{Wilcox:2021rtt,Lee:2023rmz}. 

Considering the promising implications for physics, we commence by outlining the distinctive advantages of computing the hadronic tensor through lattice QCD and point out some comparisons and its complementarity with other lattice formalisms in Sec.~\ref{sec:opportunities}. Subsequently, we will delve into the challenges associated with extracting physics information in  Minkowski space. In Sec.~\ref{sec:HT}, we briefly discuss the hadronic tensor formalism, and its implementation on the lattice is described in Sec.~\ref{sec:numerics}. We provide details of the extraction of the matrix elements  for the elastic (both using the conventional three-point function and hadronic tensor formalism) and transition form factors in Sec.~\ref{sec:extraction} and draw comparisons with the experimental data.  In Sec.~\ref{sec:prospects}, we highlight forthcoming directions, along with  potential theoretical uncertainties and limitations associated with the current method, all aimed at attaining the precision goal in future calculations and then conclude our paper.

%%%%%%%%%%%%%%%%
%%%%%%%%%%%%%%%%
%%%%%%%%%%%%%%%%

\section{Unique features and challenges in computing hadronic tensor} \label{sec:opportunities}
One significantly important aspect of the simplicity of the hadronic tensor formalism~\cite{Liu:1993cv,Liu:1999ak} is that the hadronic tensor is scale-independent and the structure functions are frame-independent. Unlike  several technical challenges associated with renormalizing the equal-time matrix elements of bilocal operators in quasi-~\cite{Ji:2013dva} and pseudo-~\cite{Radyushkin:2017cyf} PDFs approaches, which have received considerable attention in recent years~\cite{Constantinou:2017sej,Chen:2017mzz,Braun:2018brg,Ji:2020brr,LatticePartonCollaborationLPC:2021xdx}, the lattice QCD calculation of the hadronic tensor offers unique advantages in that it does not necessitate renormalization stemming from spatial Wilson line operators.

There are several lattice QCD formalisms involving two currents. Notably, a noteworthy distinction between the hadronic tensor and the spatially-separated  two-currents approach~\cite{Braun:2007wv,Ma:2014jla,Ma:2017pxb,Bali:2018spj,Sufian:2019bol,Sufian:2020vzb} lies in the fact that the hadronic tensor involves two currents that are  temporally separated. On the other hand,  the two currents' time and space are integrated in the Compton amplitude approach~\cite{Chambers:2017dov,Can:2020sxc}, and it computes the lattice matrix elements in the unphysical region, suppressing the intermediate states between the two currents from going on-shell. Conversely, the hadronic tensor in our work is evaluated in the physical region  where the intermediate states are physical.

The insertion of two currents in the evaluation of the hadronic tensor on the lattice involves an Euclidean temporal separation, which presents a major challenge: converting the lattice-calculated hadronic tensor from Euclidean space to Minkowski space. This conversion poses a formidable inverse problem, notable not only for the challenges arising from the discrete nature of data and its limited quantity but also for the intricacies involved in constructing the spectral density encompassing various elastic, inelastic scatterings, and resonance structures that collectively contribute to the hadronic tensor. However, it offers a unique opportunity to directly probe the physics of intermediate states that propagate between the currents, enabling the study of transitions from the nucleon to its various resonance states.  More detailed discussion is given in the beginning of Sec. V.

 We also note the difference in excited states associated with the nucleon three-point and four-point function calculations. In the calculation of hadronic tensors, there are two types of excited states. The first type originates from the source and sink (i.e., those between the source/sink and the currents), representing pure contamination. The second type consists of intermediate states between the two currents, which are the ones of physical interest for the transition matrix elements. These second-type excitations are not contaminants but correspond to intermediate states such as in $ N \to N $ and $ N \to X $, which are relevant for studying inclusive \( N \to X \) contributions.

Additionally, it is important to note that elastic and resonance structures are significantly suppressed in the DIS region. When analyzed within the DIS framework, the hadronic  tensor formalism provides an alternative avenue for investigating the nucleon's partonic structure. This approach complements lattice QCD calculations conducted through different methodologies, where substantial progress has already been achieved. (For many recent developments and results, see~\cite{Cichy:2018mum,Constantinou:2020hdm, Ji:2020ect,Constantinou:2022yye} and the references therein.) In principle, the hadronic tensor serves as a framework that enables the investigation of hadron structure across the elastic, resonance, and DIS regions, including the challenging transition regions that lie in between them.

\section{Calculation of elastic and transition form factors from hadronic tensor: formalism on the lattice}\label{sec:HT}

The response of the target nucleon to the photon probe is characterized by a hadronic tensor. The hadronic tensor involved in the inclusive lepton-nucleon scattering cross section is the imaginary part of the forward virtual Compton amplitude $T_{\mu\nu}$ and  can be written in the following form for spin-averaged nucleon states:
\bea \label{eq:hadtens}
W_{\mu\nu} (q^2,\nu)= \frac{1}{\pi} {\rm Im}~T_{\mu\nu} (q^2,\nu)= \frac{1}{2}\sum_n \int \prod_{i=1}^n \bigg[\frac{\dd^3 p_i}{(2\pi)^3 2 E p_i} \bigg] \bra{N(p)}J_\mu(0)\ket{n} \bra{n} J_\nu(0)\ket{N(p)}(2\pi)^3\delta^4(p_n-p-q), 
\eea
where $p$ and $p_n$ are the 4-momentum of the nucleon and $n$-th intermediate states,  $q$ is the momentum transfer ($q^2 = -Q^2$), and $\nu$ is the energy transfer. Being an inclusive reaction, the hadronic tensor in the lepton-nucleon scattering includes all intermediate states as shown in Eq.~\eqref{eq:hadtens}. $W_{\mu\nu}$ cannot be calculated in perturbation theory.  It encapsulates the aspects of nucleon structure relevant for the analysis of experimentally measured cross sections. 

 Formally, the hadronic tensor formalism expressed as in Eq.~\eqref{eq:hadtens}, can be obtained in the Euclidean space by calculating a four-point correlation of two currents on the lattice~\cite{Liu:1993cv,Liu:1999ak}. This Euclidean hadronic tensor $W^{E}_{\mu\nu}$ can be converted into the hadronic tensor in Minkowski space $W^M_{\mu\nu}$ by employing the inverse Laplace transform, wherein $\tau$ is treated as a continuous dimensionful variable: 
\bea  \label{eq:contLap_BG} 
W^M_{\mu\nu}(\vec{p},\vec{q},\nu) = \frac{E_{N,\vec{p}}}{\pi i} \int_{c-i \infty}^{c+i \infty} \dd \tau\,
e^{\nu\tau} W^{E}_{\mu\nu}( \vec{p}, \vec{q}, \tau),
\eea
with $c > 0$. However, since there is no lattice data at imaginary $\tau$, it is not a feasible approach to execute the transformation as indicated in Eq.~\eqref{eq:contLap_BG}. 
 On the other hand, this can be cast into an inverse problem 
 in the form of a Laplace transform~\cite{Liu:2016djw}: 
\bea  \label{eq:eucLap_BG}
W^E_{\mu\nu}(\vec{p}, \vec{q}, \tau) = \int_{0}^{\infty} \frac{\dd \nu}{2E_{N,\vec{p}}}~e^{-\nu \tau}W^M_{\mu\nu}(\vec{p}, \vec{q}, \nu)
\eea
In this work, we are interested in the nucleon elastic form factor and the transition form factors of the  radial excitations in the finite volume. In order to extract these form factors or structure functions, it is necessary to evaluate the nucleon two-point (2pt) and four-point (4pt) correlation functions. We can start by expressing the 2pt correlation function as follows:
\bea \label{eq:2pt}
C^{2{\rm pt}}_{\alpha\beta}(\vec{p}_f=\vec{p},t_f;\vec{p}_i=\vec{p}, t_0)=\sum_{\vec{x}_0}e^{i\vec{p}\cdot\vec{x}_0} \sum_{\vec{x}}e^{-i\vec{p}\cdot\vec{x}}\bra{0} \chi^{N}_\alpha(\vec{x},t_f)\overline{\chi}^{N}_\beta(\vec{x}_0,t_{0})\ket{0} ,
\eea
where $\chi^N(\overline{\chi}^N)$ is the nucleon  annihilation(creation) interpolation field and $\vec{p}_i,x_0,t_0$ and $\vec{p}_f,x,t_f$ are nucleon momenta, spatial, and temporal positions of the source and sink, respectively.   In the limit where $(t_f-t_0)$ is large, taking the trace for positive parity projection, we can write 
\bea
\Tr(\Gamma_e G^{2{\rm pt}})(\vec{p}_f,t_f,\vec{p}_i, t_0) \xrightarrow[]{(t_f - t_0) \gg 0} \overline{Z}_i Z_f \frac{(E_{N,\vec{p}}+m_N)}{E_{N,\vec{p}}} e^{-E_{N,\vec{p}}(t_f-t_0)},
\eea
 where $Z$ is the transition matrix element $\bra{0}\chi^N \ket{N}$ and $\Gamma_e = (1+\gamma_4)/2$. The Euclidean hadronic tensor $W^E_{\mu\nu}$ can be constructed from the following 4pt correlation function  with current insertions at two different space-time positions $(\vec{x}_2,t_{2})$ and $(\vec{0},t_{1})$ and momentum transfer $\vec{q}$ between the currents.  The nucleon 4pt correlation function can be written as:
\bea \label{eq:4pt-1}
C^{4{\rm pt}}_{\alpha\beta}(\vec{p}_f=\vec{p},t_f,\vec{q},t_2,t_1,\vec{p}_i=\vec{p},t_0)= \sum_{\vec{x}_0}e^{i\vec{p}\cdot\vec{x}_0}\sum_{\vec{x}}e^{-i\vec{p}\cdot\vec{x}}
\bra{0}\chi^{N}_\alpha(\vec{x},t_f) \sum_{\vec{x}_2} e^{-i\vec{q}\cdot \vec{x}_2} 
J_{\mu}(\vec{x}_2,t_{2})J_{\nu}(\vec{0},t_{1})\overline{\chi}^{N}_\beta(\vec{x}_0,t_{0})\ket{0} .
\eea

The current associated with the matrix element of the nucleon  form factors in the elastic region can be parametrized in terms of the Dirac $F_1$, and Pauli $F_2$, form factors: 
\bea \label{eq:vec1}
\bra{N(p)}J_{\mu}\ket{N(p')}  = \bar{u}(p)[\gamma_{\mu}F_1(-q^2) - \sigma_{\mu\nu}q_{\nu}\frac{F_2(-q^2)}{2m}] u(p'),
\eea
where $q=p-p'$, $\sigma_{\mu\nu} = \frac{1}{2i}[\gamma_{\mu}, \gamma_{\nu}]$. Here all the matrices are in Euclidean notation and we will also continue using these notations in the subsequent parts of the calculation.  On the other hand, a Lorentz decomposition of the $\gamma^*N \to N^*$ transition matrix element associated with the electromagnetic current for the  $\frac{1}{2}^+ \, N^*$ state can be written  as~\cite{Weber:1989fv,Tiator:2008kd,Aznauryan:2008us}:
\bea \label{eq:vec2}
\bra{N^*(p)}J_\mu\ket{N{(p')}} = \bar{u}_{N^*}(p)\bigg[  (\gamma_\mu-\frac{q_\mu\slashed{q}}{q^2})F^*_1(-q^2) -\frac{\sigma_{\mu\alpha}q_\alpha}{m_N+m^*}F^*_2(-q^2) \bigg ] u_N(p').
\eea
  Note that the structure of the current closely resembles its nucleonic counterpart, with the exception of the $q_\mu\slashed{q}$ component. In the case of the nucleon, the form factor related to this operator must vanish to maintain current conservation. However, this requirement does not hold for the Roper, given the difference in mass between the nucleon and Roper, with $m^* \neq m_N$.  In this work, we concentrate on the electric form factors, such as the nucleon Sachs electric form factor, $G_E$, and the nucleon to its radial excitation electric transition form factor, $G_E^*$, only. Therefore, we will explicitly choose $\mu=\nu=4$ in the above expressions for currents and matrix elements.  In this calculation, we consider the source and sink to be at rest, $\vec{p}_f=\vec{p}_i=\vec{0}$ and consider various momentum transfers at the currents. For this case, after inserting a complete set of intermediate states and with nucleon-to-nucleon (N-N) and nucleon to its first radial excitation, (N-R) scattering included, we can write Eq.~\eqref{eq:4pt-1} as the
following: 
 \begin{align}
    \label{eqn:general}
    \begin{aligned}
        \mathrm{Tr}[\Gamma_e G^{\rm 4pt}(\vec{p}_f=\vec{0},t_f,\vec{q},t_2, t_1,\vec{p}_i=\vec{0},t_0)]  & \xrightarrow[t_1\gg t_0]{t_f \gg t_2 \gg t_1}  (\Gamma_e)_{\beta \alpha}\sum_{m,n,l} \sum_{\vec{s},\vec{s}_i,\vec{s}_f}
        \bra {0} \chi_{\alpha}^{N}\ket{m,\vec{s}_f }
        \frac{m_mm_n}{E_{m,\vec{p}_f}E_{n,\vec{p}_i}}\frac{m_l}{E_{l,\vec{q}}}  \\
        &\quad \quad \quad \quad \quad \times \bra{ m,\vec{s}_f }J_{\mu}\ket{ l,\vec{q}, \vec{s} } \bra {l,\vec{q}, \vec{s}} J_{\nu}\ket{ n,\vec{s}_i }\\
        &\quad \quad \quad \quad \quad  \times 
        \bra{ n,\vec{s}_i }\overline{\chi}^{N}_\beta \ket{0} e^{- E_{n,\vec{q}}(t_2-t_1)} e^{-E_{m,\vec{p}} (t_f-t_2)-E_{n,\vec{p}} (t_1-t_0) }  \\
        &  \xrightarrow[n=m=N]{l=N,R}  
        (\Gamma_e)_{\beta \alpha} \sum_{\vec{s},\vec{s}_i,\vec{s}_f} \bra{ 0} \chi_{\alpha}^{N}\ket{ N,\vec{s}_f }
        \frac{m_N^2}{(E_{N,\vec{p}})^2}\frac{m_N}{(E_{N,\vec{q}})}  e^{-E_{N,\vec{p}} (t_f-t_0) } \\
        &\quad \quad \quad \quad \quad \times \bra{ N,\vec{s}_f } J_{\mu} \ket{ N,\vec{q}, \vec{s} } \bra{ N,\vec{q}, \vec{s} } J_{\nu} \ket{N,\vec{s}_i }\\
        &\quad \quad \quad \quad \quad  \times 
        \bra{ N,\vec{s}_i }\overline{\chi}^{N}_{\beta}\ket{0} e^{- (E_{N,\vec{q}}~ - ~ E_{N,\vec{p}})(t_2-t_1)} \\
        &\quad \quad \quad \quad +(\Gamma_e)_{\beta \alpha} \sum_{\vec{s},\vec{s}_i,\vec{s}_f} \bra{ 0 }\chi_{\alpha}^{N} \ket{ N,\vec{s}_f }
        \frac{m_N^2}{(E_{N,\vec{p}})^2}\frac{m_R}{(E_{R,\vec{q}})}  e^{-E_{N,\vec{p}} (t_f-t_0) } \\
        &\quad \quad \quad \quad \quad \times \bra{ N,\vec{s}_f }J_{\mu} \ket{ R,\vec{q}, \vec{s} } \bra {R,\vec{q}, \vec{s}} J_{\nu} \ket{N,\vec{s}_i }\\
        &\quad \quad \quad \quad \quad  \times 
        \bra{ N,\vec{s}_i } \overline{\chi}^{N}_{\beta}\ket{0} e^{- (E_{R,\vec{q}}~ - ~ E_{N,\vec{p}})(t_2-t_1)} + ...\\
        &= \overline{Z}_i Z_f   e^{-E_{N,\vec{p}} (t_f-t_0) }  \Bigg \{ \frac{E_{N,\vec{q}}+m_N}{E_{N,\vec{q}}} G_E^2  e^{- (E_{N,\vec{q}}~ - ~E_{N,\vec{p}})\tau} \\
        & + 
        \frac{E_R + m_R}{E_R} \bigg [ \frac{(E_R-m_R)(m_N+m_R)}{q^2}\bigg ]^2
        (G^*_E)^2 e^{- (E_{R,\vec{q}}~ - ~E_{N,\vec{p}})\tau}  \Bigg \} + ...,
    \end{aligned} 
\end{align}
where  $m$, $n$, and $l$ label the intermediate state with the corresponding masses $m_m, m_n$, and $m_l$, and we have written the temporal dependence of the two currents as a function of Euclidean time separation, $\tau=t_2-t_1$. Here, we have defined $G_E(Q^2)=F_1(Q^2) - \frac{Q^2}{4m_N^2}F_2(Q^2)$ for the nucleon elastic form factor. Similarly, $G_E^*(Q^2)$ can be defined from the  definition of the matrix element~\eqref{eq:vec2} as a function of $F_1^*(Q^2)$ and $F_2^*(Q^2)$ 
 as $G_E^*(Q^2) = F_1^*(Q^2) -\frac{Q^2}{(m_R+m_N)^2}F_2^*(Q^2)$. The form factors $G_E$ and $G_E^*$ can be determined from the ratio of the nucleon 4pt to 2pt functions and  we obtain
\bea \label{eqn:general3}
R^{\rm 4pt/2pt} (\vec{p}_f=\vec{0},t_f,\vec{q},t_2, t_1,\vec{p}_i=\vec{0},t_0) & \xrightarrow[t_1\gg t_0]{t_f \gg t_2 \gg t_1}& \frac{1}{2} \Bigg( {\frac{E_{N,\vec{q}}+m_N}{E_{N,\vec{q}}}} G_E^2\,  e^{- (E_{N,\vec{q}}~ - ~ m_N)\tau} \nn \\
&+& \frac{E_R + m_R}{E_R} \bigg [ \frac{(E_R-m_R)(m_N+m_R)}{q^2}\bigg ]^2
        (G^*_E)^2\, e^{- (E_{R,\vec{q}}~ - ~ m_N)\tau}  \Bigg) ,
\eea
where we have used $E_{N,\vec{0}}=m_N$. In the following, the mass $m_R$ associated with the transition matrix elements will be denoted by $m^*$ and various energy differences in Eq.~\eqref{eqn:general3} will be denoted by $\Delta E_n$.

 The charge operator $J_4$ in the present calculation contains both the isoscalar and isovector components. In principle, the isovector charge can couple the nucleon to the $\Delta$. However, it only contributes to the subdominant Coulomb quadrupole, $C_2$ form factor in the nucleon to $\Delta$ transition. The dominating magnetic dipole, $M_1$ form factor arises from the spatial charge current $J_i$~\cite{Leinweber:1992pv}. Given the partial width
of Roper (N(1440)) $\rightarrow p \gamma$ is $0.035 - 0.048\%$ and that of $\Delta \rightarrow p \gamma$ is $0.28- 0.32 \%$ and the total widths of the Roper and $\Delta$ are comparable, the photo-decay of $\Delta$ is about an order of magnitude larger than that of the Roper. However, since the $C_2/M_1$ ratio is about 5\% from a previous lattice QCD calculation~\cite{Alexandrou:2010uk} and the hadronic tensor is the square of the form factors, this
 makes  the transition to the $\Delta$ about two orders of magnitude smaller than the transition to Roper with the charge current $J_4$ that we use in this calculation. This may well be the reason that in the spectral decomposition analysis in  later sections, we don't see the $\Delta$ which is expected to be $\sim 300$ MeV above the nucleon. 

%%%%%%%%%%%%%%%%%%%

\section{Description of numerical setups}\label{sec:numerics}
\begin{figure}[htp]
	\centering
 	\includegraphics[width=6.0in, height=3.0in]{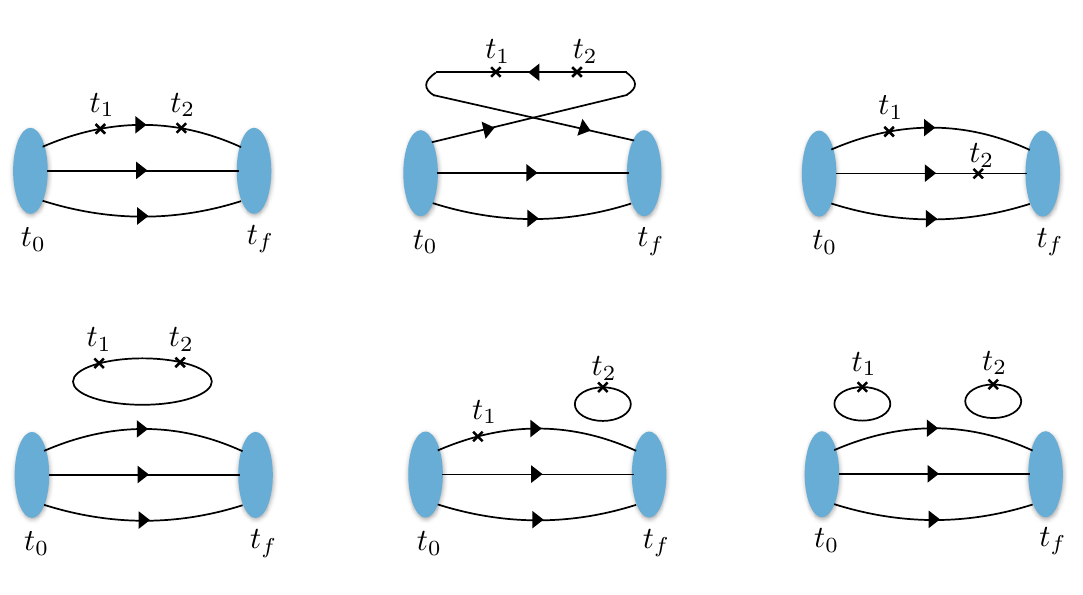}
 	\caption{Topologically distinct diagrams in the Euclidean-path integral formulation of the nucleon hadronic tensor. The flavors of the quark lines are not specified in the figure. The first  row shows the connected insertion and the  second row shows the disconnected insertion contributions. \label{fig:contractions}}
 \end{figure}

For the numerical calculation of the hadronic tensor on the lattice, one needs to calculate several topologically  distinct insertions  (both connected and disconnected) as depicted in Fig.~\ref{fig:contractions}. Different Wick contractions for the 4pt function lead to these  insertions in the Euclidean path-integral formulation. However, for this calculation, we only concentrate on the connected insertion contributions as depicted in the first row of Fig.~\ref{fig:contractions} and ignore the disconnected insertions (in the  second row of Fig.~\ref{fig:contractions}). 
In general, the Wick contraction in the third column of the  first row in Fig.~\ref{fig:contractions}, known as the cat's ear diagram in the DIS jargon describes nonperturbative interactions between the struck quark and the spectator quarks in DIS process. In the DIS, the matrix elements of the cat's ear diagram yield contributions of the structure functions that are suppressed by $1/Q^2$ and are often ignored. However, as we investigate the QE and RES regions at low energies and low $Q^2$, the contribution from this  insertion is not negligible~\cite{Liang:2020sqi} and we  have included it in this calculation.

Our calculations are conducted on the RBC/UKQCD domain wall fermion lattice, 32Ifine~\cite{RBC:2014ntl}   at a lattice spacing of $a=0.06$ fm (Table~\ref{tab:lat}). The tadpole-improved clover coefficient $c_{\rm sw} = 1.033$ is used to generate the clover term of the clover action for the valence quarks. The pion mass is tuned to be close to the unitary point $m_\pi=371$ MeV. 
%%%%%%%%%%%%%%%%%%%%%%%
\begin{table}
  \begin{center}
    \begin{ruledtabular}
      \begin{tabular}{ccccc}
      ID  & $a$ (fm)  &$m_\pi$ (MeV)  & $L^3\times N_t$ & $N_{\rm cfg}$\\
      \hline
      $32$Ifine & $a = 0.0627(3)$ & 371 & $32^3\times 64$ & 645 \\
      %\hline
      \end{tabular}
    \end{ruledtabular}
    \caption{ \label{tab:lat} Parameters for the gauge ensemble used in this work: lattice spacing, pion mass, spatial and temporal sizes, and number of configurations used.}
  \end{center}
\end{table}

We adopt similar numerical techniques from the  work in Ref.~\cite{Liang:2019frk} for the lattice QCD implementation of this calculation. We use two sequential propagators for constructing the 4pt correlation function with one starting at the nucleon source location $t_0$ through the first current at $t_1$ to the second current at $t_2$. The second one starts from $t_0$ and goes through the nucleon sink position $t_f$ to $t_2$. Therefore, for each calculation, the source point $t_0$ and the two sequential points $t_1$ and $t_f$ are fixed. This allows us to have variable $t_2$ for the second current insertion time.  For fixed hadron momentum in this calculation, three 
inversions are therefore needed for each $\vec{q}$, $t_1$ and $t_f$. 

 In contrast to the previous hadronic tensor calculation in~\cite{Liang:2019frk}, a fast fermion smearing scheme~\cite{Li:2023mmm} is used in the present work. This reduces the time cost by about 8-10 times compared to the traditional Gaussian smearing in terms of computer time consumption and is employed on both the source and sink sides to enhance the ground-state overlap. The profile of this gauge covariant fermion smearing scheme is Gaussian-like. This smearing procedure guarantees that rotation symmetry is restored after configuration averaging and has been shown to produce an effective mass of the hadron in agreement with  that using Gaussian smearing in~\cite{Li:2023mmm}. For other aspects of the calculation, we follow the numerical techniques outlined in Ref.~\cite{Liang:2019frk} for their lattice QCD implementation.

For this calculation, we choose  $t_1-t_0=4$ where $t_0$ is the source location,  and $t_f-t_0 =18, 20, 22$ in the lattice units. The second current is inserted at all temporal locations from the first current to the sink, including the positions of the first current and the sink. We therefore have $\tau = t_2-t_1$ in the  ranges $0-14$, $0-16$, and $0-18$ in lattice units. We also perform the calculation with $t_1-t_0=5$ and $6$ in lattice units to investigate the excited-state contamination from the nucleon source to the first current temporal separation.  As discussed earlier, since we are interested in the elastic and transition electric form factors, we choose $\mu=\nu=4$ for the choice of currents in Eqs.~\eqref{eq:vec1} and~\eqref{eq:vec2}. We keep the nucleon source and sinks at rest $\vec{p}_i=\vec{p}_f = \vec{0}$ and insert seven momentum transfers at the currents with $\vec{q} = [(0,0,0),(0,0,1),(0,1,1),(1,1,1),(0,0,2),(0,1,2),(1,1,2)]$ in lattice units.

 In summary, for all the calculations, we use $645$ configurations and $8$ sources  located at time slices of  $0, 8, 16, ...,56$. We also calculate 3pt functions for $6$ source-sink separations, i.e., $t_f-t_0= 10,~12,~14,~16,~18,~20$ in lattice units, and performed a joint fit for these source-sink separations and compared the results with and without excited states included in the fitting functional forms.  We calculate 4pt and 2pt correlation functions with source-sink separations $18,~20,~22$ in lattice units, with one current insertion fixed $t_1-t_0=4$. Additionally, we fix the source-sink separation at  $t_f-t_0=20$ in lattice units and vary the current insertion at $t_1-t_0=5,6,7$ to study the excited-state contamination from the nucleon source and sink. 

In the present calculation of the hadronic tensor, we do not include the disconnected insertions as they are
expected to be small compared to the connected insertions. This is true in general in the calculation of the elastic form factors in the 3-pt functions for the vector, axial, and tensor form factors. It is particularly so for the electromagnetic current due to the cancellation of the $u$, $d$, $s$, $c$ charges.
As a consequence, the disconnected insertion contributes $\sim 1\%$ to the proton magnetic moment and $\sim 2.5\%$ to the proton mean square charge radius~\cite{Sufian:2016pex,Sufian:2017osl,Alexandrou:2018zdf,Alexandrou:2018sjm,Djukanovic:2023beb,Sufian:2020coz}. Indeed, a different study of 4-point correlation functions found that for two current insertions on the same time slice, contributions from diagrams with disconnected one-current loops are much smaller than the connected contributions.~\cite{Bali:2021gel}.

\section{Matrix elements extraction for the elastic and transition form factors}\label{sec:extraction}

 One of the major systematic errors in extracting the nucleon matrix elements is  excited state contamination. The fact that there is a concomitant $\frac{1}{V}$ normalization factor associated with each hadron state in the two-point correlators has been utilized to discern the nature of a pentaquark candidate~\cite{Mathur:2004jr}  and $\sigma(600)$~\cite{Mathur:2006bs}. In the case of the nucleon correlator with $q^3$ interpolation field, it is shown that the $\pi N$ state is volume suppressed compared to the single nucleon state~\cite{Bar:2015zwa}. However, for the case
of the three-point function, there is a current-induced enhancement of the meson-nucleon state~\cite{Bar:2018xyi,Bar:2019gfx,RQCD:2019jai} due to the fact that the current can couple to a meson directly with the corresponding quantum number which propagates between the current and the source or sink. This dangling contribution with current momentum projection gains a factor of volume to compensate for the volume suppression due to the presence of the two-particle meson-nucleon state compared to the the case of a one-particle nucleon or its radially excited states. It is this realization that led to the resolution of the puzzle of the violation of the Goldberger-Treiman relation~\cite{Jang:2019vkm,RQCD:2019jai}. For the pseudoscalar and axial currents, the direct coupling to the P-wave pion state at physical pion mass and large volume produce a current enhanced low-lying P-wave $\pi N$ contamination. Upon removal of this contamination led to the recovery of the Goldberger-Treiman  relation. A study with additional $\pi N$ interpolation fields besides the conventional $q^3$ interpolator found that, at the physical pion mass,  the isovector pseudoscalar and axial matrix elements shows  significant improvement in suppression of the $\pi N$ contamination, but  this is not the case for the scalar, vector and tensor matrix elements~\cite{Alexandrou:2024tin}. This supports the current-induced enhancement picture, i.e., except for the pseudoscalar and axial current, other currents do not couple to one pion. Even though the vector and tensor currents can directly couple to the respective vector and tensor mesons, the resulting meson-nucleon states lie higher than the nucleon radial excitation such as the Roper. Thus, their contributions are exponentially suppressed.  In the present study we are dealing with vector currents. Besides, our pion mass is at 371 MeV.  Therefore, it is not a concern for the present work. In the future, when we move on to the current-current correlator with axial-vector currents at physical pion mass and large volume, we will need to take the $\pi N$ contamination involving the source and sink concurrently into account.

 In this work, we opt to fit the  4pt/2pt correlation ratio  to obtain the energies of the nucleon and excited states for each momentum transfer. We will discuss this later in the following section. In what follows, prior to employing the fit to the 4pt/2pt correlation function, we initially employ Bayesian reconstruction (BR)~\cite{Burnier:2013nla} for reconstructing the spectral functions based on the 4pt/2pt correlation. This approach allows us to investigate the qualitative features of spectral function peaks at various $\Delta E_n$ values as a consistency check.

\subsection{Bayesian Reconstruction of spectral functions}\label{sec:BR}

 The extraction of a spectral function in the cases of our lattice QCD calculations can be formulated as:
\bea
\label{eq:laplace}
    D(\tau) = \int_0^{\infty} d\omega \rho(\omega) e^{- \omega \tau},
\eea
 where $D(\tau)$ is the data that we   obtain from lattice calculations and $\rho(\omega)$ is the spectral function of the ``frequency" $\omega$, from which we can obtain physical quantities of interest,  such as the elastic and transition form factors. In this case, $D(\tau)$ and $\rho(\omega)$ are connected via the Laplace transform.

The ``time" variable, $\tau$, is generally discrete and takes $\sim\mathcal{O}(10)$  values in a limited range from the lattice calculations. In contrast, the ``frequency" variable $\omega$ is in principle continuous and takes values from  zero to infinity.  Therefore, naive extraction of spectral functions is an ill-defined ``inverse problem," where the degrees of freedom of the data are far less  than those  of the spectral function. Two directions to tackle this issue are:
\begin{enumerate}
    \item Reducing the number of degrees of the freedom in the model. In a reasonably aggressive way, it can be reduced to less than the number of degrees of the freedom in the data so the problem is no longer ill-defined. An example is the multi-exponential fit.
    \item Introducing a proper regulator with hyperparameters to the cost function, to constrain the extra degrees of the freedom of the model. An example is the Bayesian reconstruction method.
\end{enumerate}

Using the Bayesian inference, one can write the posterior as

\bea \label{eq:Bayes}
P[\rho | D, I] = \frac{P[D|\rho,I]}{P[D,I]}P[\rho|I],
\eea
where $D$ is the data, $I(\alpha,m)$ is the prior which is a function of hyperparameters $\alpha$ and the default model $m$, and $\rho$ is the model  function to be obtained. The interpretation of solving the inverse problem using the Bayesian inference perspective, is to obtain a model  function $\rho$ which yields the maximum posterior $P[\rho | D, I]$.

Lattice QCD researchers have proposed several Bayesian methods, with different expressions for the likelihood $P[D|\rho,I]$ and the prior $P[\rho|I]$ to obtain the model  spectral function $\rho$. In the context of handling the inverse problem for the  current-current correlator, the Backus-Gilbert method~\cite{BG} was introduced in~\cite{Hansen:2017mnd} and~\cite{Liang:2019frk}. Furthermore, as a Bayesian method, the maximum entropy method~\cite{Asakawa:2000tr} was also applied to solve the inverse problem in~\cite{Liang:2019frk}. It was observed that this method produced a similar outcome for reconstructing the flat spectral region, while qualitatively enhancing the reconstruction of spectral peaks when compared to the Backus-Gilbert method. Additionally, there are proposals in~\cite{Bailas:2020qmv,ExtendedTwistedMassCollaborationETMC:2022sta} for the reconstruction of smeared spectral functions from Euclidean correlation functions. For example, a formalism in~\cite{Hansen:2017mnd} was proposed to determine various transition rates from the nucleon 4pt function calculations by extracting smeared spectral functions from appropriately constructed finite-volume Euclidean correlation functions such that a well-defined infinite-volume limit exists. Reconstructing these smeared spectral functions from Euclidean correlation functions, as proposed in these prior works~\cite{Hansen:2017mnd, Bailas:2020qmv, ExtendedTwistedMassCollaborationETMC:2022sta}, is an important direction that we plan to investigate in the future. In this study, to assess the consistency of $\Delta E_n$ determination between the inversion method and exponential fitting, we will employ Bayesian reconstruction (BR)~\cite{Burnier:2013nla}. Previous lattice QCD calculations of the hadronic tensor~\cite{Liang:2019frk} have demonstrated that the BR method exhibits the highest resolution for extracting peak structures in the low-momentum transfer regions. While the BR method incorporates all the essential properties of the maximum entropy method~\cite{Asakawa:2000tr} for reconstructing positive spectral functions from lattice QCD data, it also ensures scale-invariance by enforcing that the posterior does not depend on the units of the spectral function,  leading to only ratios between the parameters of interest related to the unknown process and the default model. We briefly discuss the essential features of the BR method in the following and demonstrate its applications on the lattice data. A detailed discussion on the BR method and its comparison with Backus-Gilbert and the maximum entropy method can be found in the latest review article~\cite{Rothkopf:2022ctl}.

According to the BR method developed in~\cite{Burnier:2013nla}, the posterior to be maximized is 
\bea
\label{eq:BR_pst}
P[\rho|D,m] = \frac{P[D|\rho,I]}{P[D,m]}\int \dd\alpha P[\rho|\alpha, m], 
\eea
where the dependence on the hyperparamter $\alpha$ in the  prior  is removed by integrating over all possible values of $\alpha$ and using $P[\alpha] =1$. 

With discrete data in Euclidean time $\tau_i$ and discretization of the ``frequency" $\omega$ in the model $\rho$,  we can calculate the likelihood $P[D|\rho,I]\propto e^{-L}$ where

\bea
L = [D(\tau_i) - D^{\rho}(\tau_i)]C^{-1}_{ij} [D(\tau_j) - D^{\rho}(\tau_j)],
\eea
where $D(\tau_i)$ denotes the mean of the simulated data at the $i$th Euclidean time separation and $D^{\rho}(\tau_i)$ denotes the corresponding Euclidean data point reconstructed from the model $\rho$ based on the discretized transformation like the one in Eq. \eqref{eq:laplace}. $C_{ij}$ is the covariance matrix of the mean
\begin{equation}
    C_{ij} = \frac{1}{N_{\mathrm{conf}}(N_{\mathrm{conf}}-1) }\sum^{N_{\mathrm{conf}}}_{i=1} (D^{(k)}(\tau_i) - D(\tau_i))(D^{(k)}(\tau_j) - D(\tau_j)),
\end{equation}
where $D^{(k)}(\tau_i)$ is the $k$th measurement in the ensemble.

And the prior $P[\rho|\alpha, m] \propto e^{\alpha S'}$ can be calculated from the regulator
\bea
 S' = \sum_l \bigg[1 - \frac{\rho_l(\omega_l)}{m_l(\omega_l)}+\ln\bigg(\frac{\rho_l(\omega_l)}{m_l(\omega_l)} \bigg)\bigg]\Delta \omega_l,
\eea
where $\omega_l$ is the $l$th discrete frequency,  $\rho_l(\omega_l)$ and $m_l(\omega_l)$ are the discretized model function and default model separately, and $\Delta \omega_l$ is the size of the $l$th interval.

We can then plug the likelihood $P[D|\rho,I]\propto e^{-L}$ and prior $P[\rho|\alpha, m] \propto e^{\alpha S'}$ in Eq. \eqref{eq:BR_pst}. The corresponding posterior can be expressed as
\bea
\label{eq:BR_pst_entropy}
P[\rho|D,m] \propto \int \dd\alpha e^{S_{\rm BR}},
\eea
where
\bea
S_{\rm BR} = \alpha S' -L -\gamma(L-N_\tau)^2
\eea
Please note a Lagrangian multiplier term $\gamma(L-N_\tau)^2$ has been introduced to constrain the likelihood with condition $|L-N_{\tau}| \sim 0$ where $N_{\tau}$ is the number of data points in Euclidean time.

Given the formulas to calculate the posterior from the data $D$ and model  spectral function $\rho$, one can apply numerical methods to find $\rho$ that optimizes the posterior and take it as the solution.

\subsection{Reconstruction of spectral distributions with the BR method}

We can now map the hadronic tensor formalism to the inverse problem with a Laplace  transform, use the BR method to explore the reconstruction of the spectral weights from these 4pt/2pt correlation functions. One can extract the form factors from the spectral weights. 

Following the functional form in Eq.~\eqref{eqn:general3}, the ratio of 4pt to 2pt correlation functions can be expressed in the following functional form:
\begin{align}
    \label{eqn:general4}
    \begin{aligned}
        R^{\rm 4pt/2pt}(\tau)  &  \rightarrow \sum_{n} W_n e^{- (E_{n}- E_{N})\tau},
    \end{aligned} 
\end{align}
 where we have $W_0 \propto G_E^2$ and $W_1 \propto (G_E^*)^2$. Here we have dropped the kinematic parameters to simplify the notation. By denoting the time separation between the two inserted currents $\tau=t_2-t_1$, and comparing this with 
Eq.~\eqref{eq:laplace}, we can transform ratio of 4pt to 2pt correlation functions into a continuous form suitable for applying the BR method:
\begin{align}
    \label{eqn:BRcont}
    \begin{aligned}
        R^{\rm 4pt/2pt}(\tau)  &  \rightarrow \sum_{n} W_n e^{- (E_{n}- E_{N}) \tau} = \int d\omega \rho(\omega) e^{- \omega \tau} ,
    \end{aligned} 
\end{align}
where the spectrum to be reconstructed can be expressed as
\bea 
    \rho(\omega) = \sum_{n=0}^{N} W_n \delta(\omega-(E_n-E_{N})),
\eea
if we assume that there is no degeneracy of states in terms of energy and in our lattice data. 

Due to the discretization of the spectral function in the numerical calculations of the BR method,  the delta function will become a sharp peak around $\omega=(E_n-E_{N})$, and integration over the peak region should give an estimate of the weight $W_n$ for each discrete state $n$.

It is important to mention that the spectral functions in the BR method are twice differentiable and it gradually approximates well-defined peaks in the spectral function as the quality of the input data improves. We illustrate four such reconstructions of spectral weights and qualitative features of the various $\omega$-peaks for the currents insertion at the up-quark propagators in Fig.~\ref{fig:BR} utilizing the BR code package~\cite{Rothkopfcode}. A similar reconstructions of spectral weights of the 4pt/2pt correlation function for current insertion at the down-quark propagator is shown in Fig.~\ref{fig:BR_down}.

 We choose the default model $m$ to be constant, which is an expected result when no significant spectral information can be extracted from the data. Without specifying any prior information based on physics knowledge of the energies of the discrete states, BR lets the data tell us the values of the $E_n$'s, the energies of the intermediate states, and the corresponding spectral weights, $W_n$, by integrating over the region around the peaks. We provide examples of these reconstructed spectra from the ratio of 4pt and 2pt functions for four-momentum transfers represented as $\vec{q}=(0,0,1),(0,1,1),(1,1,1),(0,1,2)$ in lattice units. 
\vspace{0.5cm}
%%%%%%%%%%%%%%%
\befs 
\centering
\includegraphics[width=5.2in, height=3.5in]{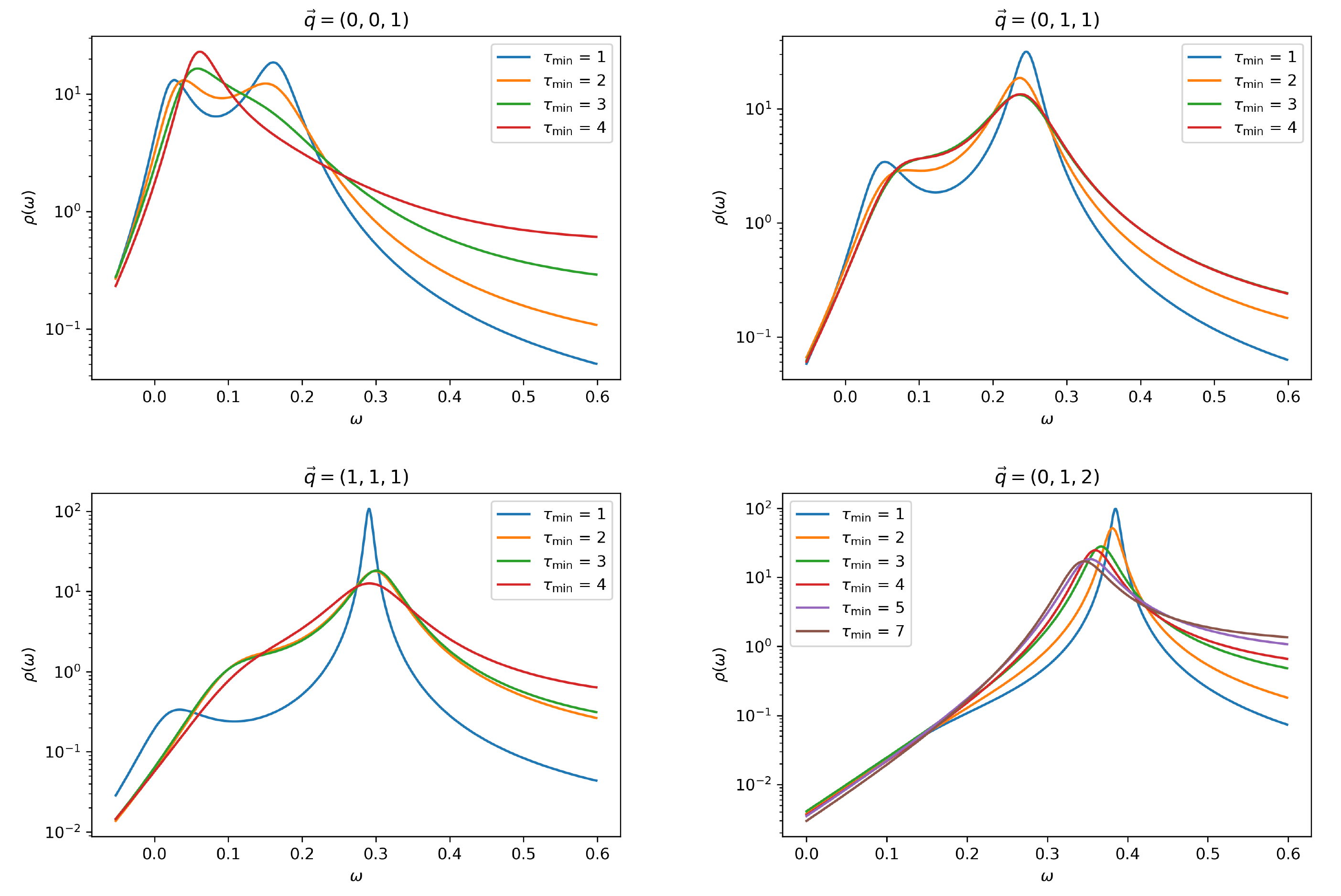}
\caption{\label{fig:BR} 
 The  reconstructed  spectral distribution  $\rho(\omega)$ obtained from the 4pt/2pt ratio using the Bayesian reconstruction method for up-quark at various momentum transfers, $\vec{q}=(0,0,1),(0,1,1),(1,1,1),(0,1,2)$. The peaks/bumps in the spectra indicate states found by the BR method from the input data, with their peak locations related to the energy of the $n$th state $\omega_{\mathrm{peak},n}\simeq E_n -E_N$. For each momentum transfer case, we also show the changes in the reconstructed spectra with input data $R^{\rm 4pt/2pt}(\tau)$ in ranges of $\tau$.} 
\eefs{mockdemocn}

\befs 
\centering
\includegraphics[width=5.2in, height=3.5in]{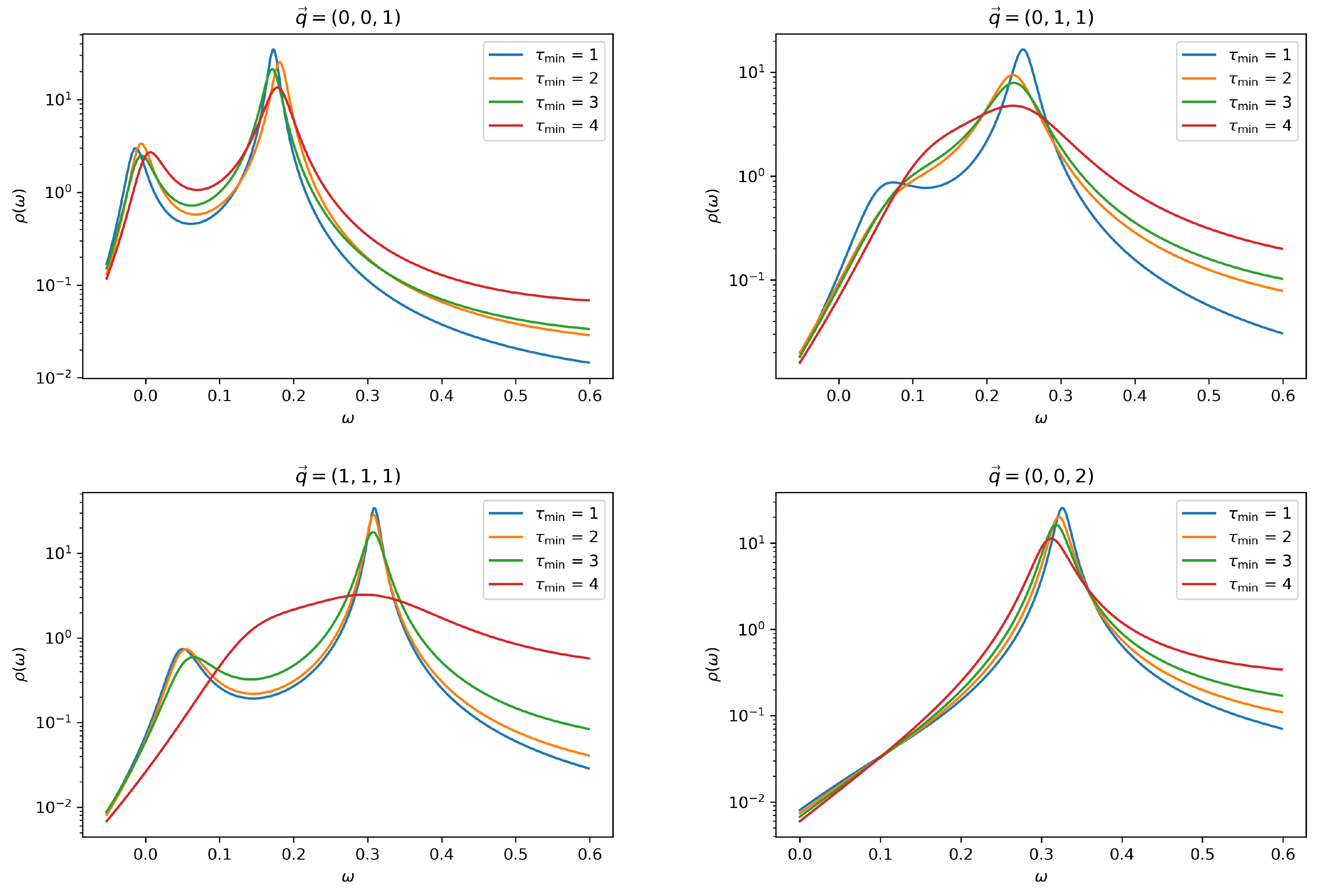}
\caption{\label{fig:BR_down} The  reconstructed spectral distribution  $\rho(\omega)$ obtained from the 4pt/2pt ratio using the Bayesian reconstruction method for down-quark at various momentum transfers, $\vec{q}=(0,0,1),(0,1,1),(1,1,1),(0,0,2)$. For each momentum transfer case, we also show the changes in the reconstructed spectra with input data $R^{\rm 4pt/2pt}(\tau)$ in ranges [$\tau_{min}$,$\tau_{max}=1,4$]. }

\eefs{mockdemocn}
%%%%%%%%%%%%%%%%%

 In Fig.~\ref{fig:BR}, we  note that, except for the largest momentum transfer case (i.e. with $\vec{q} = (0,1,2))$, there are two peak structures. Their excitation energies can be estimated from $E_n \simeq \omega_{\mathrm{peak},n} + E_N$, where $\omega_{\mathrm{peak},n}$ is the peak position of the $n$th structure in $\omega$, $E_n$ is the energy of the $n$th state and $E_N$ is the nucleon mass in our case. The energies of the first peak positions $E_0$ are close to the expected $\sqrt{m_N^2 + \vec{q}^2}$. They are naturally identified with the nucleon. For the second peaks, we estimate their invariant masses  $\sqrt{(p + q)^2} = \sqrt{m_N^2 + \omega^2 - \vec{q}^2 + 2 m_N \omega}$
for the four $|\vec{q}|$ cases, and found them to be  $1.71, 1.81, 1.92$ and 1.90 GeV, which are $0.5 - 0.7$ GeV above the nucleon mass. Since all states including those with multi-hadrons with allowed quantum numbers are accessible as intermediate states, the resonances in this mass range,  such as $1/2^+ N(1440) ({\rm Roper}), N(1710)$, and $1/2^- N(1535), N(1650)$, are likely the states that contribute to this structure. 
 As explained in Sec. \ref{sec:HT}, we do not see  the $\Delta$ at $\sim 300 $ MeV (i.e., $\omega \sim 0.1$)
above the nucleon.

 We note that, at smaller momentum transfer $|\vec{q}|$, the first peak
is prominent. As $|\vec{q}|$ increase, the peak becomes smaller and disappears at $\vec{q} = (0,1,2)$. This is because as $|\vec{q}|$ increases, so does $Q^2$ which spans the range from 0.4 to 1.9 ${\rm GeV^2}$. For the dipole form of the proton electric form factor $G_E (Q^2)$, the dipole mass squared is $M_D^2 \sim 0.71 {\rm GeV^2}$. Since the hadronic tensor for the elastic scattering is the square of $G_E (Q^2)$, its $Q^2$ dependence is, therefore, 
$1/(Q^2 + {\rm M_D^2})^4$. As a consequence, no peak is seen for the case $\vec{q} = (0,1,2)$, $Q^2 = 1.8\, {\rm GeV^2}$.
On the other hand, $Q^2= |\vec{q}|^2 - \omega^2$ ranges from $0.1 \,{\rm GeV^2}$ to $0.6\, {\rm GeV^2}$ for the second peak,  which  may include several states, are smaller as compared to the first one. This is because, given the same $|\vec{q}|$, $\omega$ is larger. 
If the states under the second structure have dipole form factors and a similar dipole mass as that of the nucleon, they will not be as suppressed as the nucleon in this range of $|\vec{q}|$. In fact, if $1/2^+ N(1440)$ and $N(1710)$ are responsible for the bulk of the structure, this $Q^2$ behavior will favor the higher state and may explain the shift of the second peak by $\sim 200$ MeV as $|\vec{q}|$ increases.

As we can see,  BR is reasonable in resolving the ground state nucleon and the  overall structure of the excited states, but not the individual excited states which might appear under the structure. Even so, it is helpful in gaining insight on the behavior of the spectral weights of the elastic peak relative to those of the excited states as a function of $Q^2$, which  is consistent with our understanding in terms of the electric form factors.  

 We also notice that, as the momentum transfer $\vec{q}$ increases, causing the elastic structure to become more suppressed, BR becomes less effective in providing accurate estimates for  $E_n$ and $W_n$ within the present statistics. 
  
 Moreover, obtaining spectral weights using the BR method requires integrations that may introduce additional systematic uncertainties. Therefore, we  shall use the conventional multi-exponential fitting for  our quantitative  results. Consequently, in the subsequent analyses aimed at determining the nucleon's elastic and resonance structures, we rely on exponential fitting as the preferred approach.

\subsection{Matrix elements extraction from fits to the correlation functions}\label{Sec:ME}

In this section, we determine the spectral weights $W_n$ and $\Delta E_n(\vec{p},\vec{q})=E_{n,\vec{q}}~ - ~ E_{N,\vec{p}}$ by fitting the 4pt/2pt correlation functions using an exponential form, as can be read from Eqs.~\eqref{eqn:general3} or~\eqref{eqn:general4}. As indicated in the BR method discussed in Sec.~\ref{sec:BR}, the current state of lattice data lacks the resolution to distinguish between  two or more closely situated peaks.   We also attempted a three-exponential fit ($n=0,1,2$) in Eq.~\eqref{eqn:general4}, but the second and third exponentials yield nearly identical fit parameters. Because the first and second excited states cannot be  separated, and the fitted parameters associated with the first and second excited states are identical within uncertainties, we perform a two-exponential fit to the data. In this case, the extracted first excited state should be regarded as an effective excited state, i.e., a linear combination of first and higher excited states.  Writing $R^{\rm 4pt/2pt} (\vec{p},\vec{q},\tau) \equiv R^{\rm 4pt/2pt} (\vec{p}_f=\vec{0},t_f,\vec{q},t_2, t_1,\vec{p}_i=0)$ for simplicity, we thus have:
\bea \label{eq:W4}
R^{\rm 4pt/2pt}_{\rm truncated} (\vec{p},\vec{q},\tau) = \sum_{n=0,1} W_n (Q^2) e^{-\Delta E_n(\vec{p},\vec{q})\tau} .
\eea
We do not impose any prior or constraint on $W_n (Q^2)$ and on the values of $\Delta E_n$. 
%\vspace{0.2cm}

As we will see from the fit parameters in the following, with the largest values of $\vec{q}=(1,1,2)$ in the lattice units, the four-momentum transfer associated with the  nucleon elastic form factor is $Q^2= 1.74(06)$ GeV$^2$, whereas for the transition form factor, the value of $Q^2= 0.48(04)$ GeV$^2$ is much smaller compared to that of the elastic structure.  These $Q^2$ values are obtained by averaging those extracted from fits to four-point correlation for the current insertions at $u$- and $d$-quark propagators. Since the two flavors are fitted separately, the resulting $Q^2$ values differ slightly, but they are consistent within uncertainties, as can be seen from the flavor-separated form factors shown in Fig.~\ref{fig:GEFF}.

%%%%%%%%%%%%%%%
\befs 
\centering
\includegraphics[scale=0.65]{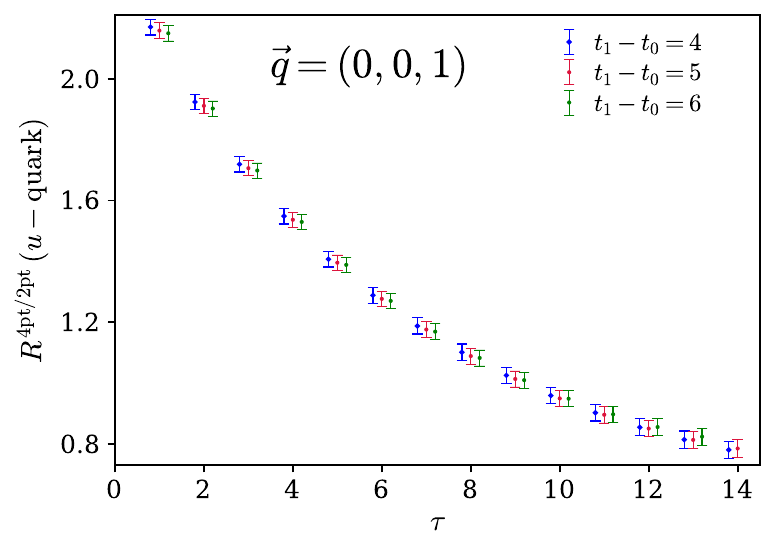}
\includegraphics[scale=0.65]{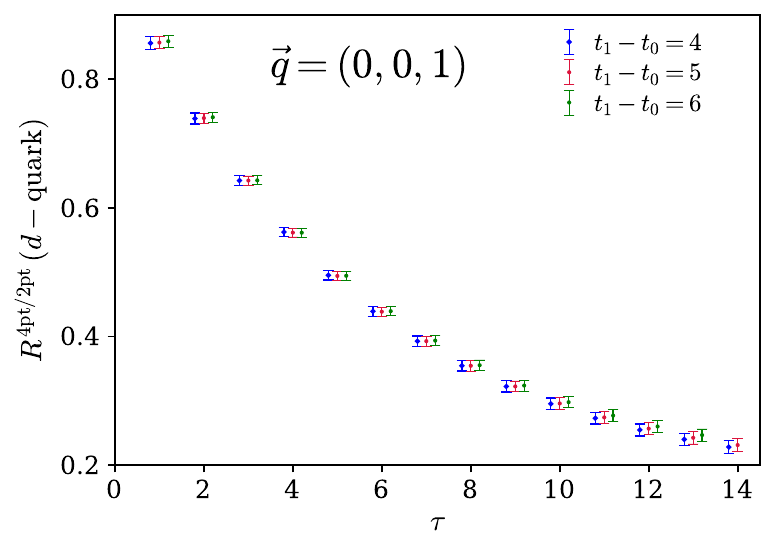}

\includegraphics[scale=0.65]{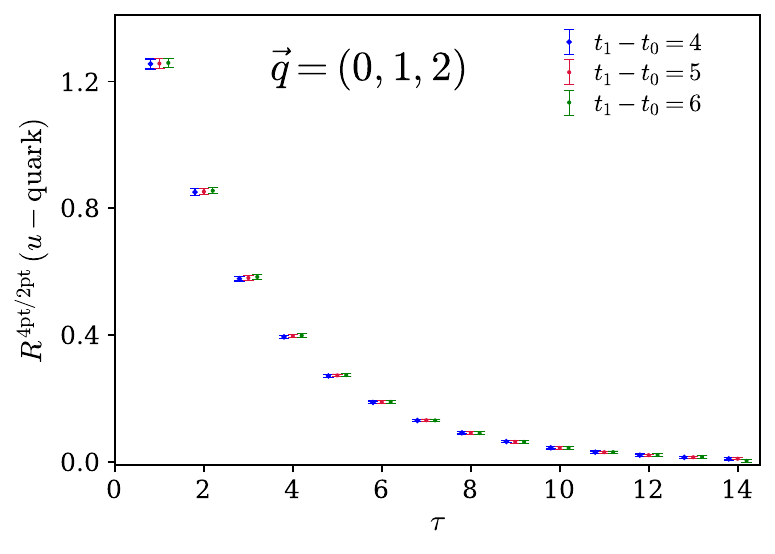}
\includegraphics[scale=0.65]{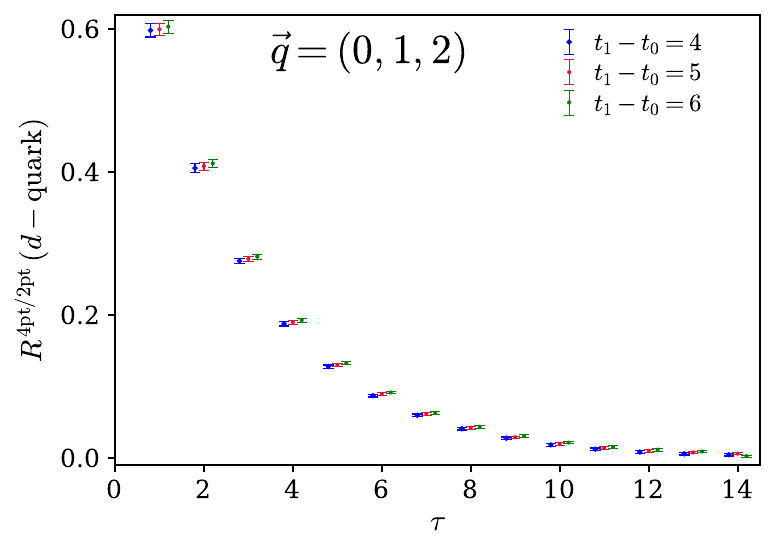}

\includegraphics[scale=0.65]{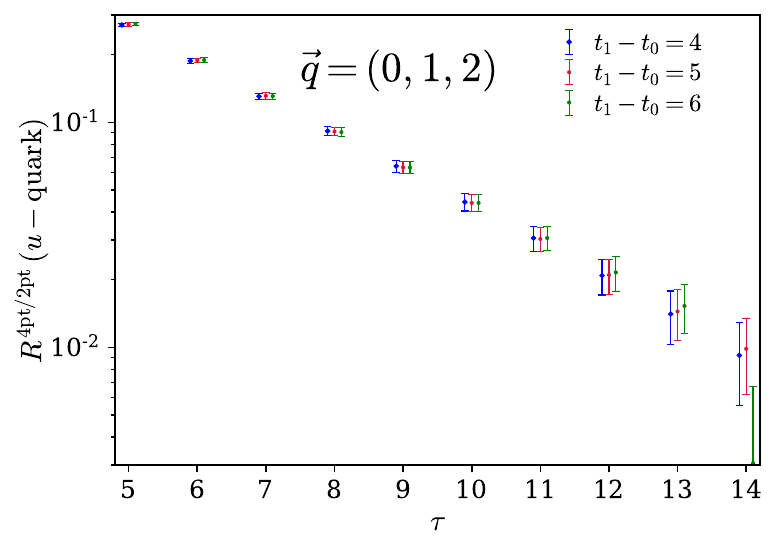}
\includegraphics[scale=0.65]{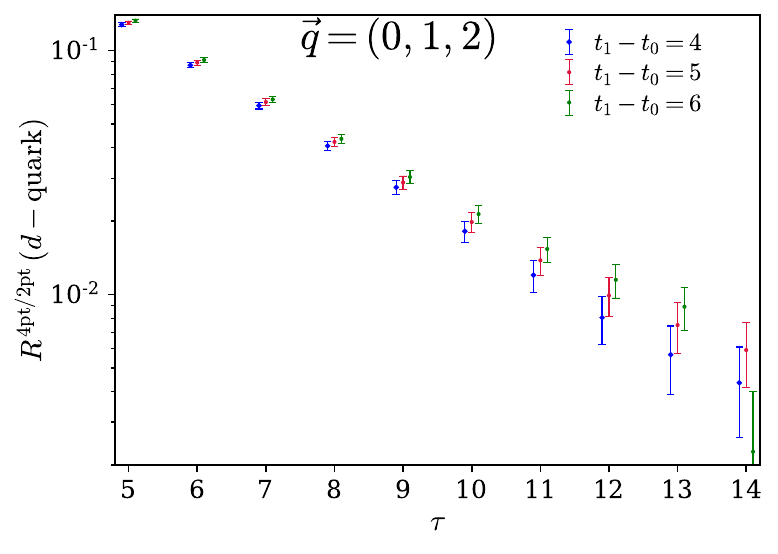}
\caption{\label{fig:t1dependence} 
Investigation of excited-state contamination as a function of the nucleon source to the first current temporal separation $t_1$ for $\vec{q}=(0,0, 1)$ and $(0,1, 2)$. Three different $t_1-t_0=4,5,6$ for which the maximum values of $\tau$ are $\tau_{\rm max}=16,15,14$ have been chosen for comparison.  The $t_1-t_0=4$ and $t_1-t_0=6$ data points  are
shifted by $0.2$  to the left and right for better visibility.   The left-upper panel figure shows such a comparison for the up-quark correlation function, labeled by $R^{\rm 4pt/2pt} \,{ (u-\rm quark)}$ for $\vec{q}=(0,0, 1)$ and the right-upper panel shows such a comparison for the down-quark correlation, $R^{\rm 4pt/2pt} \,{ (d-\rm quark)}$. The second-row figures show such a comparison for $\vec{q}=(0,1, 2)$. For a better illustration, in the third row, we zoom in on the data points for the $u$ and $d$-quark matrix elements for $\vec{q}=(0,1,2)$ in the range $\tau=5-14$  and plot the $y$-axis in logarithmic scale. } 
\eefs{mockdemocn}
%%%%%%%%%

As discussed in Sec.~\ref{sec:HT}, we have chosen to fix the temporal position of the first current at  $t_1-t_0=4$ in  lattice unit. However, it is essential to ascertain whether the potential contamination arising from the excited states, originating from the temporal separation between the nucleon source and the first current at $t_1$, remains sufficiently small. This determination is crucial to justify our decision to maintain $t_1$ at a specific value for subsequent analysis. To address this concern, we calculate the matrix elements at two different momentum transfers, $\vec{q}=(0,0,1)$ and $(0,1,2)$, and investigate the level of excited-state contamination as a function of temporal separations between the nucleon source and the first current, specifically by varying the position of the first current at $t_1-t_0=4,5,6$  both for the current insertions at the $u$- and $d$-quark propagators. 

 The results for the three $t_1-t_0$ are shown in 
Fig.~\ref{fig:t1dependence}. By virtue of the fact that there are no discernible differences within errors in these three cases, it illustrates that the excited-state contamination stemming from the nucleon source to the first current is nearly negligible.   For a better illustration, in the third row of Fig.~\ref{fig:t1dependence}, we zoom in on the data points for the $u$- and $d$-quark matrix elements for the  momentum transfer $\vec{q}=(0,1,2)$ in the range $\tau=5-14$. The agreement among data points for different values of $t_1-t_0$ indicates that excited-state contamination has minimal dependence on the temporal separation between the nucleon source and the first current insertion for a fixed source-sink separation of the nucleon. Consequently, this validates our choice of $t_1-t_0=4$ for use in subsequent analyses.

We show  two  examples of exponential fits according to the fit Eq.~\eqref{eq:W4} in Fig.~\ref{fig:Mefffit} below for the determination of the form factors.  In order to demonstrate that the excited state contamination is not sensitive to the source sink-separation for this calculation, in the second row of Fig.~\ref{fig:Mefffit}, we zoom in on the data points for the range $\tau=4-15$ and show that within error bars, the data points are in very good agreement.

To be specific, the fit form in Eq.~\eqref{eq:W4} is 
\bea \label{eq:W2}
R^{\rm 4pt/2pt}_{\rm truncated} (\vec{p},\vec{q},\tau) =  W_0 (Q^2) e^{-\Delta E_0(\vec{p},\vec{q})\tau}+W_1 (Q^2) e^{-\Delta E_1(\vec{p},\vec{q})\tau}
\eea
%
%%%%%%%%%%%%%%%
\befs 
\centering
\includegraphics[scale=0.6]{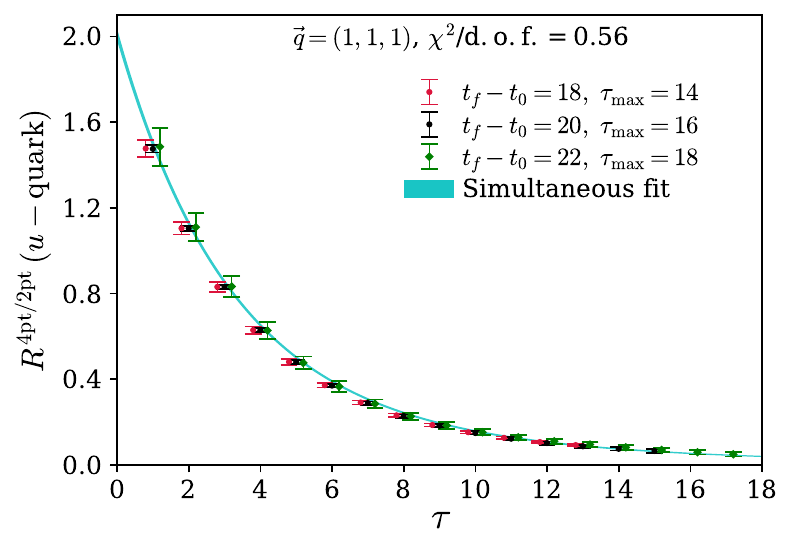}
\includegraphics[scale=0.6]{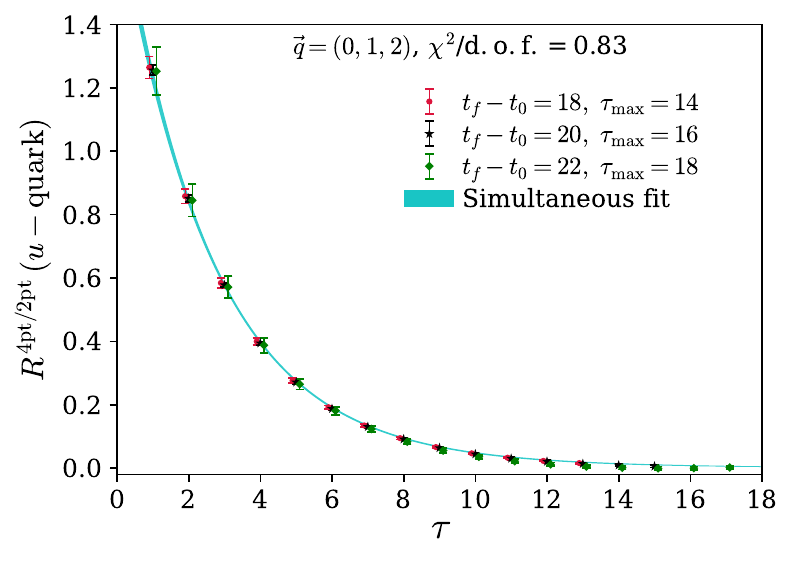}

\includegraphics[scale=0.6]{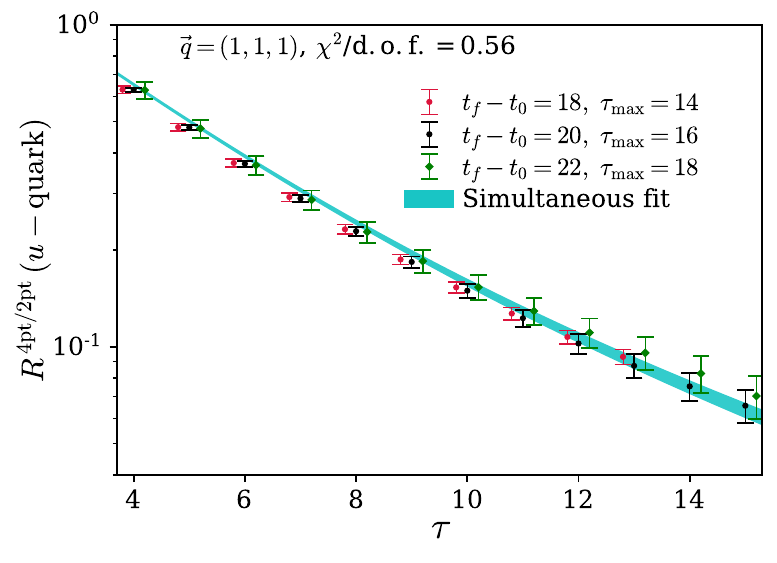}
\includegraphics[scale=0.6]{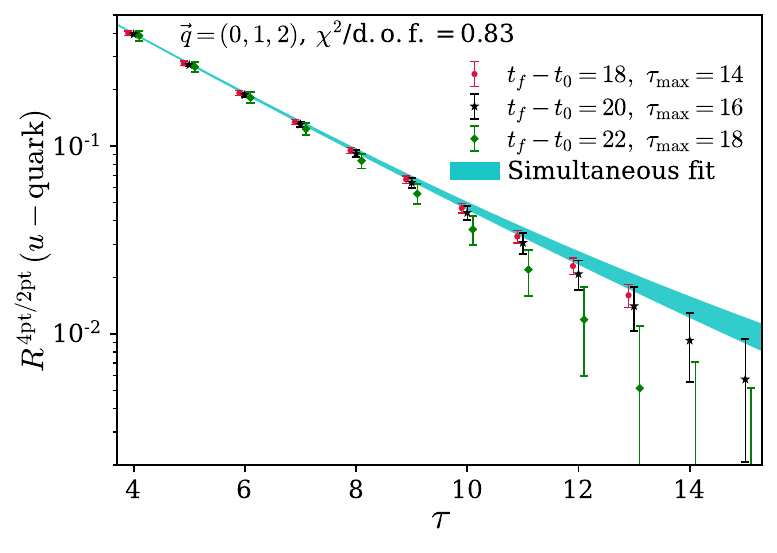}

\caption{\label{fig:Mefffit} 
 Simultaneous extraction of spectral weights $W_n$ and energy gap $\Delta E_n$ between the ground-state nucleon and its excited states due to momentum transfer  from datasets for $\tau_{\rm max}=14~(t_1-t_0=4,~t_f-t_0=18,~\tau=t_2-t_1=0-14),~\tau_{\rm max}=16~(t_1-t_0=4,~t_f-t_0=20,~\tau=t_2-t_1=0-16),~\tau_{\rm max}=18~(t_1-t_0=4,~t_f-t_0=22,~\tau=t_2-t_1=0-18)$. The relative distance $t_1-t_0=4$ is kept fixed and the second current is inserted at all temporal locations from the first current to the sink, including the positions of the first current and the sink. The datasets for $\tau_{\rm max}=14$ and $\tau_{\rm max}=18$ are
shifted by $0.2$ to the left and right for better visibility.  The figure in the left-upper panel shows fit to the up-quark 4pt/2pt correlation function $R^{\rm 4pt/2pt} \,{ (u-\rm quark)}$ for  three-momentum transfer $\vec{q}=(1,1,1)$. The cyan bands show the fitted band constructed from the parameters using fit Eq.~\eqref{eq:W2}. The figure in the right-upper panel shows fit to the  up-quark 4pt/2pt correlation function for three-momentum transfer $\vec{q}=(1,1,2)$ in the lattice units.   For demonstration purposes, in the second row, we zoom in on the same data points of the plots in the first row for the range $\tau=4-15$ and the $y$-axis is plotted in logarithmic scale. }     
\eefs{mockdemocn}
%%%%%%%%%%%%%%%%%

 We perform a correlated simultaneous fit to the $\tau_{\rm max}=14,16,18$ datasets using the functional form in Eq.~\eqref{eq:W2}. The purpose of choosing three different source-sink separations $t_f-t_0=18, 20, 22$ is to investigate the contamination coming from the nucleon's sink side to the second current. Additionally, since these three different source-sink separations also result in three different  largest temporal separations between the two currents, i.e. $t_2-t_1 = \tau_{\rm max} = 14, 16, 18$, this analysis also helps us investigate the dependence of the intermediate states between the two currents.  

 For all momentum transfers, the matrix elements for the datasets with $\tau_{\rm max}=14, 16, 18$ are fitted in the ranges $\tau = 3\text{–}12$, $3\text{–}14$, and $3\text{–}15$, respectively, for both up- and down-quark matrix elements. The choice of $\tau \geq 3$ is made to avoid $\chi^2/{\rm d.o.f.} > 1.3$ for a small subset of matrix elements and to ensure that the second current is not too close to the temporal location of the first current. The exclusion of the largest $\tau$ values is motivated by their increasing statistical noise, as well as to prevent the second current from being too close to the nucleon sink time.  As pointed out earlier, we have tried to fit $R^{\rm 4pt/2pt}_{\rm truncated} (\vec{p},\vec{q},\tau)$ in Eq.~\eqref{eq:W4} with $n=0,1,2$,  but we find that the second and third states are degenerate with the same energy and spectral weights. In other words, one cannot distinguish them and there is no information about any higher states from the data. This is the same as we learned from the BR in the last section. Thus, we shall report with two-state fits, i.e., $n$ is up to 1. In Table~\ref{tab:fitexp}, we list the nucleon  and the excited state invariant masses extracted from fitting $R^{\rm 4pt/2pt}$ across different momentum transfers.

%%%%%%%%%%%%%%%%
\begin{table*}
  \centering
  \setlength{\tabcolsep}{5pt}
  \renewcommand{\arraystretch}{1.9}
  \begin{tabular}{cccc}
  \toprule
    $\vec{q}=(q_x,q_y,q_z)$ &  $m_0=m_N$ (GeV) & $m_1^*$ (GeV) & Average $\chi^2/{\rm d.o.f.}$\\
    \midrule
    (0,0,1) & $1.26(07)$ & $1.98(05)$  & 1.1 \\
    (0,1,1) & $1.25(04)$ & $1.97(04)$  & 0.58\\
    (1,1,1) & $1.31(04)$ & $2.00(03)$  & 0.88\\
    (0,0,2) & $1.19(12)$ & $2.02(03)$  & 0.72 \\
    (0,1,2) & $1.28(14)$ & $2.15(04)$  & 0.92\\
    (1,1,2) & $1.22(11)$ & $2.06(04)$  &  0.71\\
    \bottomrule
  \end{tabular}
\caption{Nucleon and its radial excitation masses extracted from the fits to the 4pt/2pt correlation function using the functional form in Eq.~\eqref{eq:W2}. The $\Delta E_n$ values are converted to masses in the unit of GeV.   The average $\chi^2/{\rm d.o.f.}$ reported in the fourth column is obtained by performing separate fits to the up- and down-quark matrix elements, and then averaging the resulting $\chi^2/{\rm d.o.f.}$ values.}\label{tab:fitexp}
\end{table*}
%%%%%

\subsection{Elastic form factor}
Various components of the hadronic tensor, when computed on the lattice, provide valuable insights into different nucleon properties. This includes the determination of the electromagnetic and axial form factors, as well as insights into nucleon dynamics through response functions or structure functions. In this lattice QCD calculation, one of the primary objectives is to determine the nucleon form factors by evaluating the hadronic tensor in both the elastic and resonance regions, which are relevant for  lepton-nucleon scattering studies.   In this first lattice QCD calculation of the hadronic tensor in the elastic and resonance regions, it is important to make a  comparison between the elastic form factors obtained from the traditional nucleon three-point function calculation and those from the hadronic tensor approach. At this stage, this comparison is for a demonstration of the qualitative correctness and feasibility of the hadronic tensor method. This will help lay the groundwork for a future high-statistics determination of elastic and transition form factors from a single lattice QCD calculation. While the 3pt function method remains the preferred technique for precisely determining nucleon electromagnetic and axial form factors with lower computational cost than 4pt function calculations, the hadronic tensor formalism offers a significant advantage when exploring multi-hadron states, as well as shallow and deep inelastic regions. In these cases, the hadronic tensor naturally includes inclusive contributions, making it a  viable framework for such studies~\cite{Kronfeld:2019nfb}. This serves as the primary motivation for performing the hadronic tensor calculation.  In this article, our focus is primarily on the electric form factor, and we plan to address the magnetic and axial form factors in a forthcoming work, while also delving into the investigation of the nucleon-to-Delta transition form factors. 

In this section, we compare the flavor-separated nucleon Sachs electric form factor determined from the conventional nucleon three-point function calculation to that obtained from the hadronic tensor. First, we briefly mention the $G_E(Q^2)$ calculation from the traditional approach of the nucleon 3pt function calculation. We determine flavor-separated $G_E(Q^2)$ for the first four-momentum transfers (including the zero momentum transfer) to compare with those determined from the hadronic tensor calculation as a consistency check of the hadronic tensor formalism. In the computation of the nucleon 3pt function, we do not include the disconnected $u/d$-quarks insertions  as their contributions were found to be much smaller in previous calculations~\cite{Sufian:2017osl,Alexandrou:2018zdf,Alexandrou:2018sjm,Djukanovic:2023beb}. 

For the nucleon, with the source at rest, the ratio of the 3pt to the 2pt function can be used for the determination of $G_E(Q^2)$. Following Ref.~\cite{Wang:2020nbf}, the functional form with first excited state terms can be written as: 
\begin{align}
\label{eq:ratio_sqrt}
    \begin{aligned}
        R^{\rm 3pt/2pt}(t,\tau;\vec{p}_i=\vec{0},\vec{p}_f) & = \frac{ C_{N,3pt}^{S_iS_f}(t,\tau;\vec{p}_i=\vec{0},\vec{p}_f)}{C_{N,2pt}^{S_f}(t;\vec{p}_f)} \sqrt{\frac{C_{N,2pt}^{S_i}(t-\tau;\vec{p}_i)C_{N,2pt}^{S_f}(t;\vec{p}_f)C_{N,2pt}^{S_f}(\tau;\vec{p}_f)}{C_{N,2pt}^{S_f}(t-\tau;\vec{p}_f)C_{N,2pt}^{S_i}(t;\vec{p}_i)C_{N,2pt}^{S_i}(\tau;\vec{p}_i)}} \\ 
        &\approx G_{E}(Q^2) \sqrt{\frac{m_N + E_{\vec{p}_f}}{2E_{\vec{p}_f}}} + C'_1 e^{-\Delta E^1_i \tau} +  C'_2 e^{-\Delta E^1_f(t-\tau)} + C'_3 e^{-\Delta  E^1_i \tau - \Delta E^1_f(t-\tau)}, 
    \end{aligned}
\end{align}
where, $\vec{p}_i$ and $\vec{p}_f$ are the source and sink momenta and $\vec{q} = \vec{p}_f-\vec{p}_i$ is the three-momentum injected by the current, and $\tau$ is the current insertion time. We perform 3pt calculations for 6 source-sink separations: $t=10, 12, 14, 16, 18, 20$ in lattice units.  We fit the ratios in Eq.~\eqref{eq:ratio_sqrt} in two ways: the constant fit and the reduced ratio fit with first excited state terms (Eq.~\eqref{eq:ratio_sqrt}) for multiple source-sink separations.  Similar to the method used in Ref.~\cite{He:2021yvm}, we use a Gaussian prior term in addition to the conventional $\chi^2 \equiv \sum_{ij}(D_i-\tilde{D}_i)C^{-1}_{ij}(D_j-\tilde{D}_j)$ to stabilize the fitting of the ground state by constraining the excited states~\cite{Lepage:2001ym}. We introduced the above-mentioned Gaussian priors for the excited state energies using the energy differences of the nucleon mass and the first excited states ($\Delta E_i$'s and $\Delta E_f$'s) obtained from the 2pt fits. On the other hand, we have studied the consistency of the extracted spectrum and matrix elements from various analyses, especially the dependence of the results on the priors, including the case where we only use the mean values of $\Delta E_i$'s and $\Delta E_f$'s as the initial values of fitting, i.e., no priors. We found the results from the ground state are robust against the choices of the priors.

We present examples of the ratios and the fitted results in Figs. \ref{fig:GE_3pt_REM_up} and \ref{fig:GE_3pt_C1_up} separately for these two fitting procedures. The results exhibit consistency among themselves, reflected in a goodness-of-fit statistic of $\chi^2/{\rm d.o.f} \sim 1$. To facilitate a  comparison, we incorporate the fitted values obtained from the constant fit at $t=18$. We calculate the differences for the constant fits at $t=14, 16$, summing them in quadrature to estimate the systematic uncertainty, as illustrated in Fig.~\ref{fig:GEFF}. Since the excited-state fit results have smaller uncertainties and are mostly affected by smaller source-sink separation data,  due to the lack of ability to pin down the excited state, we choose the constant fit at $t=18$ which has an error  larger than the excited-state fit. However, the 3pt function calculation here is for demonstration purposes only and we do not invest more computing time to  a precise estimation of these 3pt function calculations. A precise and high statistics determination of the nucleon elastic form factor with proper estimation of systematic uncertainty using 3pt function is an active research area in lattice QCD~\cite{Park:2021ypf,Djukanovic:2022wru,Alexandrou:2023qbg,Tsuji:2023llh,Djukanovic:2024krw}, which is not the focus of our calculation.  

\befs 
\centering
\includegraphics[width=5.5in, height=2.0in]{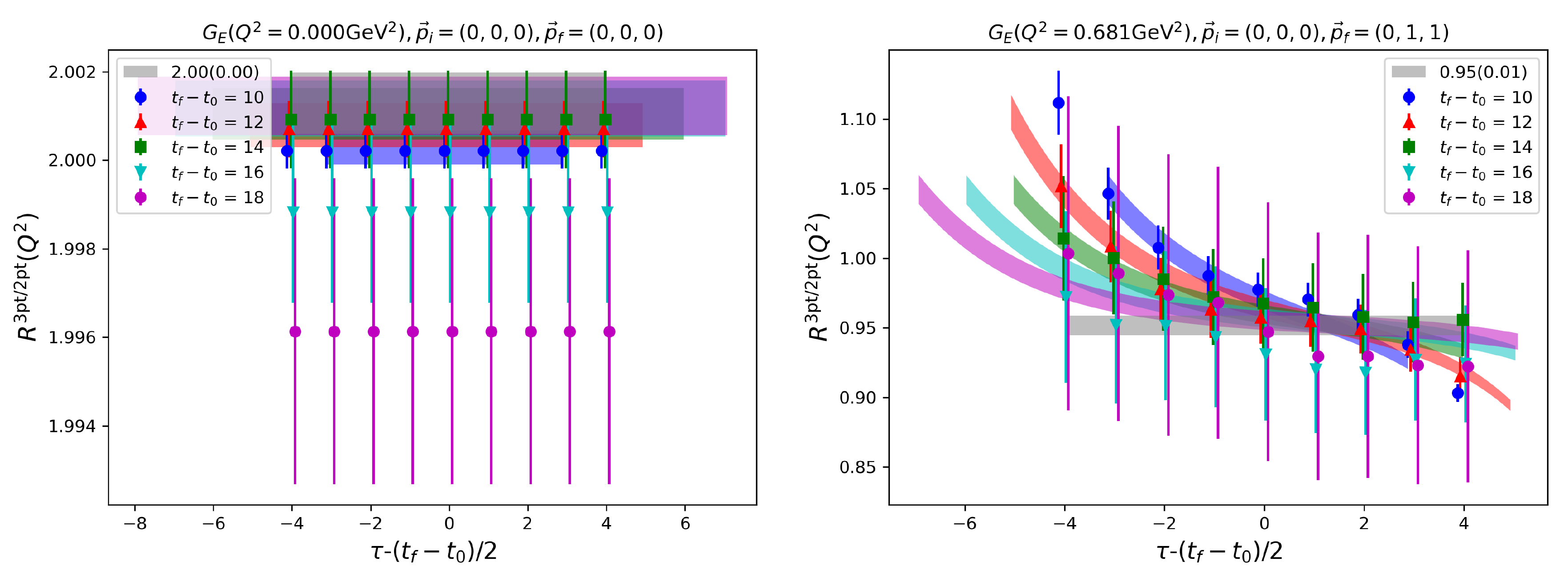}
\caption{\label{fig:GE_3pt_REM_up} Examples of the $G_E^u$ form factor fitted with the ratio method with the first excited-state terms included for the up quark. $R^{\mathrm{3pt}/\mathrm{2pt}}(Q^2)$ denotes the ratio constructed from 3pt and 2pt functions defined in Eq. \eqref{eq:ratio_sqrt}. The colored bands are reconstructed ratios using the fitting results for each separation and the grey band is the fitted form factor.  The left panel is for the forward case with no momentum transfer and the right panel is with momentum transfer $q = (0, 1, 1)$.}      
\eefs{mockdemocn}
%%%%%%%%%%%%%%%%%
%%%%%%%%%%%%%%%

\befs 
\centering
\includegraphics[width=5.5in, height=2.0in]{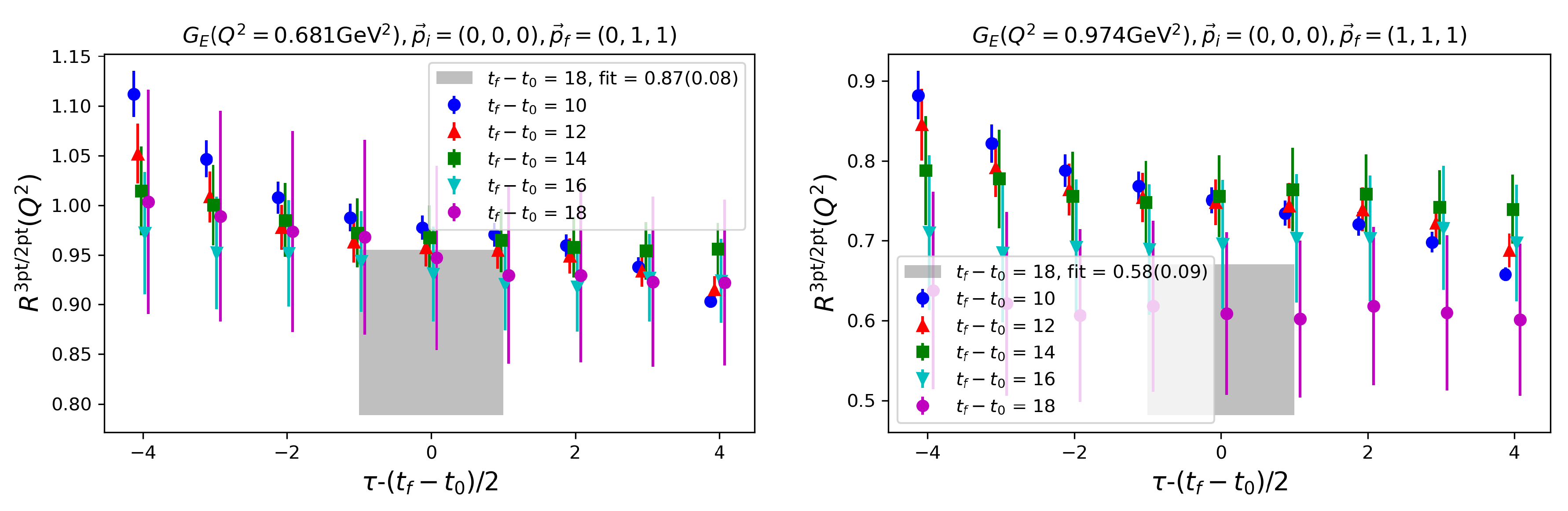}
\caption{\label{fig:GE_3pt_C1_up} Examples of the $G_E^u$ form factor fitted with constant fit for the up quark at source-sink separation  18 in the lattice units. $R^{\mathrm{3pt}/\mathrm{2pt}}(Q^2)$ denotes the ratio constructed from 3pt and 2pt functions defined in Eq. \eqref{eq:ratio_sqrt}. The grey bands represent the fitted form factors.}  
\eefs{mockdemocn}

As mentioned earlier, we have incorporated the differences between the fitting values at source-sink separation $t=18$ and those at $t=14, 16$ to estimate the systematic uncertainty in the determination of the $G_E$ form factor through the 3pt function calculation. It is worth noting that the fit to the  matrix elements at $t=18$ for $Q^2=0.68$ and $0.98$ GeV$^2$ exhibit larger uncertainties compared to those at $t=14, 16$. This can be seen from the $G_E^u(Q^2)$ at $Q^2=0.68$ GeV$^2$, determined from $t=18$ data points as illustrated using the fit band in Fig.~\ref{fig:GE_3pt_C1_up}. Along with these large uncertainties coming from the fit to the matrix elements at $t=18$, the differences from the fits at $t=14, 16$ are included in the systematic uncertainties and are reflected as large error bars in the $G_E(Q^2)$ at $Q^2=0.68$ and $0.98$ GeV$^2$ in Fig.~\ref{fig:GEFF}. However, as can be seen from the right panel in Fig.~\ref{fig:GE_3pt_C1_up}, the matrix elements at $t=18$ are still statistically consistent with the lattice data points at $t=14, 16$ which have much smaller uncertainties. Given that we applied a constant fit to the $t=18$ data points due to our inability to perform a proper excited-state fit to the 3pt functions, as a conservative approach, we present these constant fits  at $t=18$ with a larger uncertainty as our final results.

Next, we discuss the determination of the nucleon elastic form factor $G_E(Q^2)$ using the fit described in Sec.~\ref{Sec:ME} using the hadronic tensor formalism.  It is known theoretically that the hadronic tensor for the elastic scattering (i.e., with the nucleon as the  intermediate state between the currents) is the product of the elastic form factors for the two currents. Since we use the charge current in this study, we can immediately check
the vector Ward identity by verifying that the forward limit of the charge-charge hadronic tensor
is the charge squared for the separate valence quarks, i.e., $(2 e_u)^2$ for the $u$ quark and $e_d^2$ for the $d$ quark. This result was published in~\cite{Liang:2020sqi} and we also show the results in Fig.~\ref{fig:testiq0}.
\befs 
\centering
\includegraphics[scale=0.7]{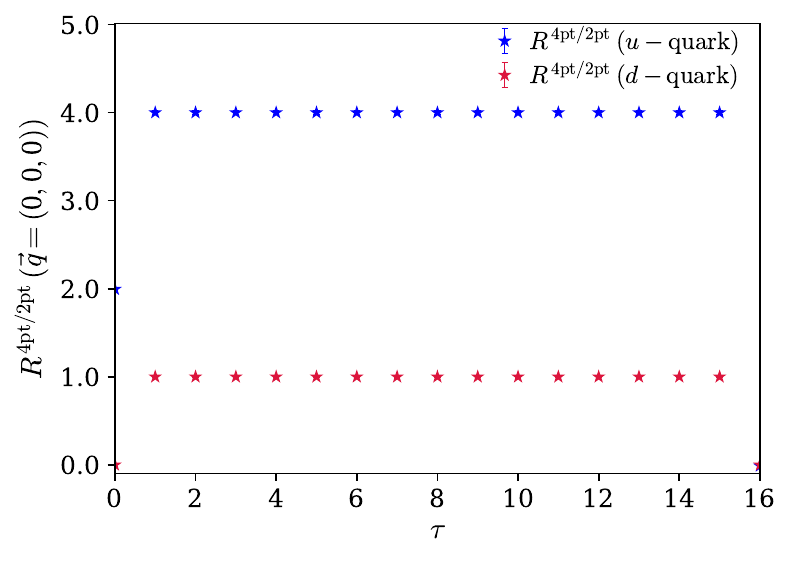}
\caption{\label{fig:testiq0} 
Square of the total charges of the $u$ and $d$ quarks in the proton in the forward limit, normalized by the square of the individual quark charges.}   \eefs{mockdemocn}

The $G_E(Q^2)$ form factor can be extracted from the fitted parameter of the spectral weight $W_1$  by fitting the matrix element using the fit form~\eqref{eq:W2}. All the kinematic factors  are obtained from the fit parameter $\Delta E_1$ for a given $\vec{q}$. In Fig.~\ref{fig:GEFF}, we show the determination of the   $G_E^{u,d}(Q^2)$ form factor for the $u$ and $d$ quarks obtained from the hadronic tensor in the range of $Q^2 \in [0,1.74]$ GeV$^2$  with a combined fit to $\tau_{\rm max}=14,16,18$ datasets.  Note that, the four-momentum transfers, $Q^2$ in the hadronic tensor formalism are obtained from the fitted parameters $\Delta E_n$ and that is why they have uncertainty along the $x$-axis in Fig.~\ref{fig:GEFF}.
 
%%%%%%%%%%%%%%%
\befs 
\centering
\includegraphics[scale=0.7]{Fig9.pdf}
\caption{\label{fig:GEFF} 
 Nucleon Sachs electric form factor $G_E^{u,d}(Q^2)$ for the $u$ and $d$ quarks in the range of $Q^2 \in [0,1.74]$ GeV$^2$ using the hadronic tensor formalism. For comparison purposes, $G_E^{u,d}(Q^2)$ form factors obtained from the nucleon 3pt function are also plotted for the first three nonzero momentum transfers as well as  $Q^2=0$ GeV$^2$. The error bars in the determination of $G_E(Q^2)$ from the 3pt function encompass both statistical and systematic uncertainties, which are added in quadrature. }      
\eefs{mockdemocn}
%%%%%%%%%%%%%%%%%

 We present a comparison of  $G_E^{u,d}(Q^2)$ determined from the 4pt and 3pt functions in Fig.~\ref{fig:GEFF}. It is important to note that the $Q^2$-values obtained for the nucleon 3pt function calculation derived from the nucleon mass and the three-momentum transfer $\vec{q}$ through the dispersion relation may not necessarily coincide with those obtained in the 4pt functional calculation. However, they are found to be consistent within the uncertainty range of the $Q^2$-values obtained through the hadronic tensor formalism.  The numerical agreement observed in the $G^{u,d}_E(Q^2)$ values between the 3pt and the 4pt extractions affirms the  feasibility of extracting form factors using the hadronic tensor formalism and our method to extract it using the fit form in Eq.~\eqref{eq:W2}.

\subsection{Nucleon to its excited states transition form factor}\label{sec:GEstar}
In this section, we present the results of the nucleon's transition to its finite-volume excitations from the $W_1$ spectral weights, as determined according to Eq.~\eqref{eq:W2}  with two terms. All kinematic factors are determined from the fitted parameters $\Delta E_n$. Given that we employed the fit form described in Eq.~\eqref{eq:W2} for fitting the hadronic tensor, we tried to determine the properties of both the first and the second finite-volume radial excitations of the nucleon. However, as previously discussed, the current  statistics of the lattice data do not allow distinguishing two closely situated states, where the separation is on the order of approximately $300$ MeV. Consequently, 
 we report results  with the understanding that the obtained excitation energy represents the mixed average of those of   $1/2^+ N(1440) ({\rm Roper}), N(1710)$, and $1/2^- N^*(1535), N(1650)$ and a suppressed contributions from $\Delta$ as discussed above.  The obtained spectral weight represents the mixed average of the squared form factors 
of these resonances. 
Keeping this in mind, we shall nevertheless compare with the transition form factor $G_E^*(Q^2)$ between the nucleon and the Roper resonance.

%%%%%%%%%%%%%%%
\befs 
\centering
\includegraphics[scale=0.7]{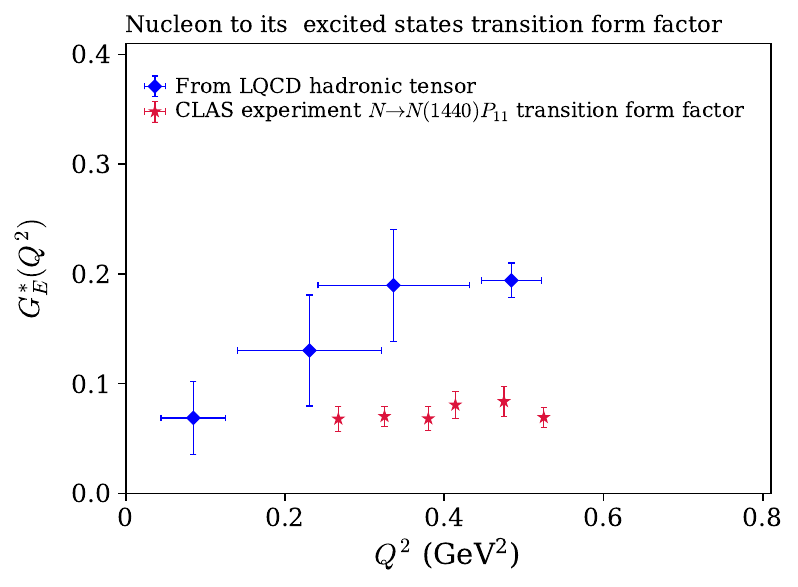}
\caption{\label{fig:GErho2} 
 Determination of the nucleon to its excited states transition electric form factor in the momentum transfer range of $Q^2 \in [0.08,0.48]$ GeV$^2$. Based on the discussion in the main text, this state is qualitatively identified with the Roper state in comparison with the  CLAS experimental data~\cite{CLAS:2009ces,CLAS:2012wxw} of $N\to N(1440)P_{11}$ transition form factor. The experimental data points~\cite{CLAS:2009ces,CLAS:2012wxw} are compiled from Ref.~\cite{Mokeev:2023zhq} only in a limited range of $Q^2$ for comparison with the lattice QCD calculation within the hadronic tensor formalism.}      
\eefs{mockdemocn}
%%%%%%%%%%%%%%%%%

We first note that for $\vec{q}=(0,0,1)$ and $(0,1,1)$ the four-momentum transfers $Q^2$ determined from $\Delta E_n$  are negative.  In Fig.~\ref{fig:GErho2}, 
we plot our lattice QCD determination of $G_E^*(Q^2)$ for $Q^2>0$ and the CLAS experimental results for the Roper~\cite{CLAS:2009ces,CLAS:2012wxw,Burkert:2017djo, Mokeev:2023zhq}.  We see from the figure that the lattice result with the smallest $Q^2 \approx 0.08\, {\rm GeV^2}$ is about the same size as the experimental data at higher $Q^2$; whereas, the errors  of the lattice results at these higher $Q^2$ are large, so that one cannot draw definitive conclusions at this stage. Our results are consistent with the earlier lattice calculation of the nucleon to Roper transition with the 3-pt function approach, which also has large errors~\cite{Lin:2008qv}. We should also keep in mind the fact that our lattice is small and the pion mass is heavy.  Moreover, the second state will contain $N(1710)$, $N(1650)$ and a very small contribution from isospin 3/2 intermediate states in addition to the Roper.

 Besides the transition form factor, we shall examine the helicity amplitude to gain a different perspective.
Through the nucleon electroexcitation reactions, $eN \to e'N^*$, the  excitation of nucleon resonances occurs through the intermediate processes $\gamma^* N \to N^*$, where $\gamma^*$ is a virtual photon, and the cross sections of the processes can usually be expressed in terms of the electromagnetic transition form factors, or helicity amplitudes $S_{1/2}(Q^2)$ and $A_{1/2}(Q^2)$.  Since we have extracted only the $G_E^*$ form factor,  the relevant amplitude is the longitudinal helicity amplitude, $S_{1/2}(Q^2)$, which  measures the resonance response to the longitudinal component of the virtual photon:
\bea \label{eq:s12}
S_{1/2}(Q^2) = \sqrt{2\pi\alpha\frac{Q^2+(m_2^*-m_N)^2}{m_N(m_2^{*2}-m_N^2)}}\frac{m_2^*+m_N}{2\sqrt{2}\,Q^2\,m_2^*}\sqrt{[Q^2+(m_2^*-m_N)^2][Q^2+(m_2^*+m_N)^2]}~G_E^*(Q^2),
\eea
where $\alpha \approx 1/137$ is the fine structure constant.

 In Fig.~\ref{fig:S12}, we compare the helicity amplitude $S_{1/2}(Q^2)$ from Eq.~\eqref{eq:s12}, determined from the hadronic tensor calculation for the excited state(s) with the corresponding values extracted from the CLAS Collaboration experimental data.

 We should point out  that the presence of $Q^2$ in the denominator of $S_{1/2}$ in Eq.~\eqref{eq:s12} results in large uncertainties in the lattice data points, especially when the values of $Q^2$ are small in the present calculation. Nonetheless, the data presented in Fig.~\ref{fig:S12} underscores the  potential  impact of the lattice QCD calculations in determining the helicity amplitudes at very small $Q^2$, particularly near the real photon point where the longitudinal helicity amplitude $S_{1/2}$ is maximal. This contribution from lattice QCD calculations is particularly valuable in complementing the low $Q^2 \to 0$ region where there are not many data points available from experiments.
%%%%%%%%%%%%%%%
\befs 
\centering
\includegraphics[scale=0.7]{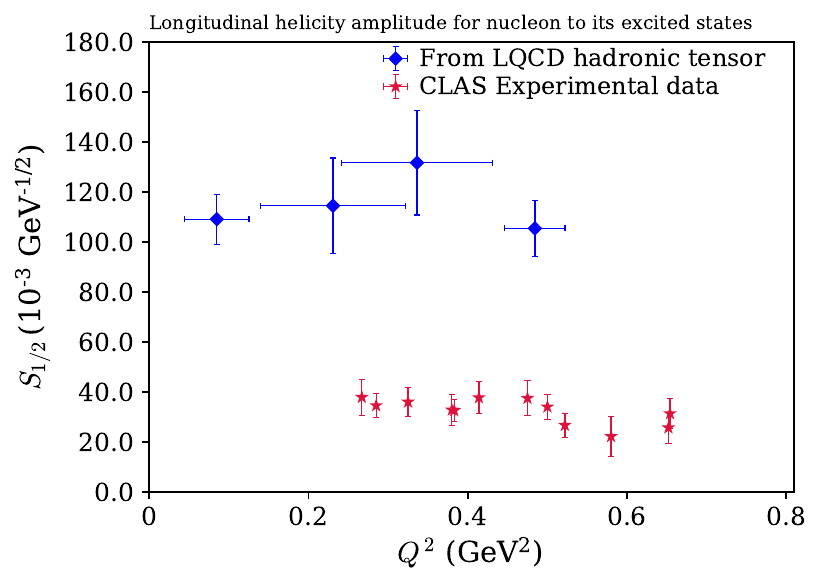}
\caption{\label{fig:S12} 
 Comparison of the longitudinal helicity amplitude, $S_{1/2}(Q^2)$ associated with the lattice QCD extraction of the nucleon to its excitation transitions compared with the experimental extraction of the $S_{1/2}(Q^2)$ associated with nucleon to Roper transition by the CLAS collaboration~\cite{CLAS:2009ces,CLAS:2012wxw}. The estimation of the $S_{1/2}$ amplitude in lattice QCD is based on the $G_E^*(Q^2)$ form factor, as detailed in the main text. The red data points represent the experimental extraction of $S_{1/2}$ by the CLAS collaboration~\cite{CLAS:2009ces,CLAS:2012wxw}, compiled from the reference~\cite{Mokeev:2023zhq}. }      
\eefs{mockdemocn}
%%%%%%%%%%%%%%%%%

We see that the lattice results are within a factor of 3
compared to the experimental data for Roper in the range $Q^2 \approx 0.08 - 0.48\, {\rm GeV^2}$.  This could be due to the fact that the lattice results have included $N(1710)$  and $N^*(1650)$, it could also be due to the fact that our present lattice is small and the pion mass is high. This should be addressed in future calculations with large lattices and physical pion mass. It would be interesting to check the partial-wave-reversal effect for the $1/2^+$ and $1/2^-$ states with the $J_i$ currents.

\section{Discussions and conclusions}\label{sec:prospects}

In this paper, we present the results of a lattice QCD investigation into the determination of the nucleon's elastic and  inelastic scatterings  through a single calculation, employing the hadronic tensor formalism for the first time.  We have checked that the electric form factors $G_E(Q^2)$ for both the $u$ and $d$ quarks in elastic scattering agree with those from the traditional 3pt function calculation.  For the inelastic scattering with the charge current $J_4$, we observe an excitation structure whose invariant mass is at 0.5 -- 0.7 GeV above the nucleon  in the BR analysis. Since the hadronic tensor is inclusive, which includes all intermediate states with allowed quantum numbers,  it could include $N(1440) (\rm Roper)$, $N(1710)$, $N^*(1535)$ and a small contribution from isospin 3/2 states in this mass range. It could also include multi-hadron states such as $\pi N$ and $\pi \pi N$.  Note these multi-hadrons states are physical states between the currents. They are different from the $\pi N$ contamination between the source/sink and the currents.    Neither the BR inverse algorithm, nor a multi-exponential fit can differentiate these states in the structure. We compare the transition electric form factor from the exponential fit of the calculated hadronic tensor with the experimental $G_E^*(Q^2)$ and the longitudinal helicity amplitude $S_{1/2}(Q^2)$ of the Roper state. We find the lattice $S_{1/2}(Q^2)$, albeit with large errors, is within a factor of three from the Roper data in the range $Q^2 = 0.08 - 0.48\, {\rm GeV^2}$. 

The intermediate states we considered, i.e., $N(1440), N(1710)$ and $N^*(1535)$ are resonances in reality. When the pion mass is near its physical value, these finite volume discrete states will be mixed with $\pi N$ states and be shifted 
depending on the volume. The phase shift is obtained from the L\"{u}scher quantization condition~\cite{Luscher:1986pf}. To match to the infinite volume Minkowski hadronic tensor, the corresponding Euclidean version has a finite volume Lelloch-L\"{u}scher correction factor~\cite{Lellouch:2000pv} which can be applied to the
intermediate states before taking the infinite volume limit. The more general application is given in Ref.~\cite{Briceno:2019opb}.

 We should emphasize that although the inverse algorithm and the exponential fitting are not able to resolve individual states, the hadronic tensor includes all the resonances and the multi-hadron states. When an isolated structure is identified in BR such as in the present study, one can take the integrated spectral weight in the support of the energy range and present the total lepton-nucleon scattering cross section in that energy bin to confront experiments.

While this study represents a preliminary exploration in this direction, the semi-quantitative agreement observed with the experimental data is both promising and motivating, despite  the fact that the lattice is small and the pion mass is not at the physical point. It underscores the prospective utility of the future development of the hadronic tensor formalism in understanding nucleon's low-lying resonance structures and in offering crucial constraints on the physics involved in modeling many-body nuclear theory, especially in the context of studying neutrino oscillation experiments. In particular, the hadronic tensor formalism includes the inclusive contribution of all the intermediate states which is crucial to providing information for the neutrino scattering experiments at low energies. Furthermore, the hadronic tensor formalism can be extended to the investigation of nucleon's deep inelastic structure functions. This feature enhances the comprehensiveness of our approach, enabling the simultaneous examination of the nucleon's elastic, resonance, and deep inelastic structures within a unified framework.

\section{Acknowledgement}
We would like to thank all the members of the $\chi$QCD collaboration for fruitful and stimulating exchanges. We are grateful to A. Rothkoph for letting us use his code for the Bayesian Reconstruction. K.F.L. wishes to thank S. Brodsky, H. Meyer, and J.W. Qiu  for fruitful discussions. R.S.S. acknowledges A. Hanlon, T. Izubuchi, L. Leskovec, A. Meyer, S. Mukherjee, and V. Mokeev for helpful discussions. This work is  supported in part by the Office of Science of the U.S. Department of Energy under Grant No. DE-SC0013065 and No. DE-AC05-06OR23177, which is
within the framework of the TMD Topical Collaboration.  
J. L. is supported by the National Natural Science Foundation of China (NSFC) under Grant Nos.\ 12222503 and 12175073.
R.S.S. was supported by the Special Postdoctoral Researchers Program of RIKEN and RIKEN-BNL Research Center. R.S.S. is also supported by Laboratory Directed Research and Development (LDRD No. 23-051) of BNL
and RIKEN-BNL Research Center. C. Z. is supported by the Alexander von Humboldt Foundation. The authors also acknowledge partial support by the U.S. Department of Energy, Office of Science, Office of Nuclear Physics under the umbrella of the Quark-Gluon Tomography (QGT) Topical Collaboration with Award DE-SC0023646. This work used Stampede time under the Extreme Science and Engineering Discovery Environment (XSEDE), which is supported by National Science
Foundation Grant No. ACI-1053575. We also used resources on Frontera at Texas Advanced Computing Center (TACC).  We also thank the National Energy Research Scientific Computing Center (NERSC)
for providing HPC resources that have contributed to the
research results reported within this paper. We acknowledge the facilities of the USQCD Collaboration used for
this research in part, which are funded by the Office of
Science of the U.S. Department of Energy. 

%%%%%
%%%%%
%%%%%
%\clearpage
\bibliography{References.bib}
%\clearpage

\end{document}